\documentclass{aastex631}

\shorttitle{Commensal Searches in SGRB and SNe Fields}
\shortauthors{Chastain et al.}

\graphicspath{{./}{figures/}}
\usepackage{rotating}

\begin{document}

\title{Commensal Transient Searches with MeerKAT in Gamma-Ray Burst and Supernova Fields}

\correspondingauthor{S. I. Chastain}
\email{sarchast@ttu.edu}
\author{S. I. Chastain}
\affiliation{Department of Physics and Astronomy, University of New Mexico, 210 Yale Blvd NE, Albuquerque, NM, 87106, USA}
\affiliation{Department of Physics, George Washington University, 725 21st St NW, Washington, DC, 20052, USA}

\author{A. J. van der Horst}
\affiliation{Department of Physics, George Washington University, 725 21st St NW, Washington, DC, 20052, USA}

\author{A. Horesh}
\affiliation{Racah Institute of Physics, The Hebrew University of Jerusalem, Jerusalem 91904, Israel}

\author{A. Rowlinson}
\affiliation{Anton Pannekoek Institute for Astronomy, University of Amsterdam, Postbus 94249, 1090 GE Amsterdam, The Netherlands}
\affiliation{ASTRON, the Netherlands Institute for Radio Astronomy, Oude Hoogeveensedijk 4, 7991 PD, Dwingeloo, The Netherlands}

\author{A. Andersson}
\affiliation{Astrophysics, Department of Physics, University of Oxford, Keble Road, Oxford, OX1 3RH, UK}

\author{R. Diretse}
\affiliation{The Inter-University Institute for Data Intensive Astronomy (IDIA), University of Cape Town, Private Bag X3, Rondebosch 7701, South Africa}

\author{M. Vaccari}
\affiliation{Inter-University Institute for Data Intensive Astronomy, Department of Astronomy, University of Cape Town, 7701 Rondebosch, Cape Town, South Africa}
\affiliation{Inter-University Institute for Data Intensive Astronomy, Department of Physics and Astronomy, University of the Western Cape, 7535 Bellville, Cape Town, South Africa}
\affiliation{INAF - Istituto di Radioastronomia, via Gobetti 101, 40129 Bologna, Italy}

\author{R. P. Fender}
\affiliation{Astrophysics, Department of Physics, University of Oxford, Keble Road, Oxford, OX1 3RH, UK}

\author{P. A. Woudt}
\affiliation{Department of Astronomy, University of Cape Town, Private Bag X3, Rondebosch 7701, South Africa}

\begin{abstract}
		The sensitivity and field of view of the MeerKAT radio telescope provides excellent opportunities for commensal transient searches. We carry out a commensal transient search in supernova and short gamma-ray burst fields using methodologies established in~\citet{commensal1}. We search for transients in MeerKAT L-band images with 30 minute integration times, finding 13 variable sources. We compare these sources to the VLASS and RACS survey data, and examine possible explanations for the variability. Additionally, for one of these sources we examine archival \textit{Chandra} ACIS data. We find that 12 of these sources are consistent with variability due to interstellar scintillation. The remaining source could possibly have some intrinsic variability. We also split the MeerKAT L-band into an upper and lower half, and search for transients in images with an 8 second integration time. We find a source with a duration of 8 to 16 seconds that is highly polarized at the lowest frequencies. This source is spatially coincident with a star detected by the Transiting Exoplanet Survey Satellite (\textit{TESS}). We conclude that this source may be consistent with a stellar flare. Finally, we calculate accurate upper and lower limits on the transient rate using transient simulations. 
\end{abstract}

%\keywords{radio continuum: transients -- stars: flare -- quasars: general }
	%%%%%%%%%%%%%%%%%%%%%%%%%%%%%%%%%%%%%%%%%%%%%%%%%%

%%%%%%%%%%%%%%%%% BODY OF PAPER %%%%%%%%%%%%%%%%%%

\section{Introduction}
\label{sec:introduction}
With the establishment of new radio telescopes such as MeerKAT~\citep{2009IEEEP..97.1522J}, ASKAP~\citep{2008ExA....22..151J}, and LOFAR~\citep{2013A&A...556A...2V}, large areas of the sky are being imaged in a relatively short period of time. This has led to surveys such as the LOFAR Multi-frequency Snapshot Survey \citep[MSSS;][]{2016MNRAS.456.2321S}, the LOFAR Two-metre Sky Survey (LoTSS) \cite{2017A&A...598A.104S}, the ASKAP Variables and Slow Transients Survey \citep[VAST;][]{2021PASA...38...54M}, the Rapid ASKAP Continuum Survey \citep[RACS;][]{2020PASA...37...48M}, the Caltech-NRAO Stripe 82 Survey \citep[CNSS;][]{2016ApJ...818..105M}, and the VLA Sky Survey \citep[VLASS;][]{2020PASP..132c5001L}. These surveys are finding a great variety of transients and variables on a wide range of timescales. 

Gamma-ray burst (GRB) afterglows are transients that, at radio wavelengths, evolve on timescales of days up to years for the brightest events. \cite{2021MNRAS.503.1847L} searched for radio afterglows in RACS observations, finding a radio afterglow candidate for GRB 171205A. Tidal disruption events (TDEs) are another class of transient that is visible in radio and evolves on timescales of days to years. In the Caltech-NRAO Stripe 82 Survey (CNSS), \cite{2020ApJ...903..116A} found a TDE. This was a very early finding that showed the promise of finding TDEs in large radio surveys. \cite{2024arXiv240608371D} found 12 TDE candidates in the VAST Pilot Survey. \cite{2024ApJ...974..241A} found a number of TDEs in the RACS survey, constraining detection rates, emission timescales, and the fraction of radio bright TDEs. \cite{2025ApJ...982..163S} present a sample of TDEs detected and selected from VLASS radio observations. Two of these events that are particularly interesting are discussed in \cite{2023arXiv231003795S}.  In VLASS, \citet{2023ApJ...948..119D} discovered a rare radio transient that would have been very difficult to find in targeted observations. \cite{2015ApJ...806..224M} discusses the predicted transient rates for GRBs, TDEs, as well as SNe and potential magnetar emission in these surveys, and as these surveys continue will be an interesting point of reference. All the aforementioned transients would historically be observed in targeted observations. However, since new radio surveys re-observe the same field multiple times over long timescales and have excellent sky coverage, they are enabling for searches for populations of rare, bright transients that would normally require a large number of individual pointings that would result in many non-detections. 

On shorter timescales of minutes to seconds, a number of new transients are being discovered, whose astrophysical origins are not yet well understood. \cite{10.1093/mnras/stv2797} found a non-repeating transient in the MSSS with a duration of less then eleven minutes. A new, growing class of radio transients on these timescales are long-period repeating transients (LPRTs) with high linear polarization. \cite{2024NatAs.tmp..107C} found a mode-switching transient with a 54 minute period within a target of opportunity observation of GRB 221009A. \cite{2023Natur.619..487H} discovered a radio transient with a 21 minute period as part of a program with the Murchison Widefield Array \citep[MWA;][]{2013PASA...30....7T} to monitor the galactic plane. Recent observations of these transients have started to uncover the astrophysical origin of these events. \cite{2024ApJ...976L..21H} report on a long-period radio transient with an optical counterpart corresponding to M-dwarf star. \cite{2024arXiv240811536D} associate one of these transients with a M-dwarf - white dwarf binary system. These transient detections have been made in searches of images and have been enabled by the excellent instantaneous (u,v)-coverage and sensitivity, and large field of view, of these new radio telescopes. 

In addition to the aforementioned transients, commensal transient searches in MeerKAT data have yielded interesting results as well: the first MeerKAT transient was found in the same field as the black hole X-ray binary GX 339-4~\citep{2020MNRAS.491..560D}, along with several variable sources \citep{2022MNRAS.512.5037D}; and several variables were found in the same field as the black hole X-ray binary MAXI J1820+070 \citep{2022MNRAS.517.2894R}. In addition, \cite{commensal1} recently presented the results of commensal searches for transients on multiple timescales in eight short GRB fields. In this search, a large number of variables was found, of which 120 were variable due to interstellar scintillation and 2 showed possible intrinsic variability. All these sources were variable on time scales of days to months, and searches at shorter timescales, 15 minutes and 8 seconds, did not result in significant transient detections. Interstellar scintillation is a phenomenon caused by the radio waves from a distant, typically compact source, interacting with charged particles in the interstellar medium. This interaction causes the source to appear to dim and brighten. Interstellar scintillation can be informative about some of the properties of the interstellar medium and the size of the radio source. 

In the study presented here, we perform commensal searches in fields of supernovae (SNe) and short GRBs observed by MeerKAT. These fields have been observed for 4 to 5 hours at a time, which allows for transient searches between different epochs but also at shorter timescales. We follow the methodology of \citet{commensal1}, in particular the software and techniques established in that study, by making images with approximately 30-minute integration times. This allows us to probe the transient parameter space at timescales of hours, but also on timescales of days to years by comparing multiple observations of the same field. We also made images with 8-second integration times of the full bandwidth, and of the lower and upper half of the bandwidth. Given the large fractional bandwidth of the MeerKAT L-band ($\Delta\nu/\nu\approx 0.6$), splitting up the band allows us to to potentially capture narrow-band or steep-spectrum emission that would be missed by imaging the full bandwidth. Since the loss of sensitivity scales as approximately the square root of the bandwidth, the loss of sensitivity is not large compared to the potential scientific gain. Additionally, splitting the band allows for keeping data that would otherwise be affected by radio frequency interference (RFI), since we can still have the other half of the band. In this paper, when it comes to the 8-second timescales, we will focus on the split-band images. 

In Section~\ref{sec:observation3}, we describe the observations used in the survey and data processing strategies. In Section~\ref{sec:methods3}, we describe the techniques and software used to find transient sources in the images. Then, in Section~\ref{sec:results3}, we give an overview of the interesting sources that came out of our search. This is followed by Section~\ref{sec:discussion3}, in which we discuss the possible counterparts and identification of the sources we find, along with the calculated transient rate as a result of this survey. Finally, in Section~\ref{sec:conclusion3} we draw overall conclusions based on this study.

\section{Observations}
\label{sec:observation3}

Deep observations were taken of short GRB and SN fields, and the sources in the field center were examined in detail in \cite{2024ApJ...970..139S}, \cite{2024MNRAS.532.2820C}, and Ruiz-Carmona et al. (in prep.). Each observation was approximately 4 to 5 hours in length, and the entire set of observations is summarized in Table~\ref{tab:allobs}. During the observations, the pointing of the telescope cycled between the target for approximately 15 minutes and a complex gain calibrator for about 5 minutes. In addition, a bandpass and flux calibrator was observed at either the beginning or end of the observations for approximately 10 minutes. These calibrators are also listed for each observation in Table~\ref{tab:allobs}. 
\begin{sidewaystable}

%\begin{table*}
	\caption{Observations used in this study, with the start and end times (in UT), phase center position, time spent on the target, and calibrators used.}
	\label{tab:allobs}
	\begin{tabular}{lllllll}
		\hline
		Name &                               Observation Start \& End Time &       RA (deg) &      Dec (deg) &  Time (hrs) &   Bandpass & Complex Gain \\
		\hline
		SN2019np & 2019-01-11 23:15:47.5 $-$ 2019-01-12 03:56:56.4 & 157.3415 &  29.5107 &                  3.22 & J1331+3030 &   J1120+1420 \\
		SN2019np &  2019-03-10 19:50:15.6 $-$ 2019-03-11 00:13:8.9 & 157.3415 &  29.5107 &                  2.97 & J1331+3030 &   J1120+1420 \\
		SN2019muj & 2019-08-09 00:09:48.9 $-$ 2019-08-09 05:08:41.4 &  36.5771 &  -9.8359 &                  4.11 & J0408-6545 &   J0240-2309 \\
		SN2020ue & 2020-01-14 01:43:23.7 $-$ 2020-01-14 06:12:48.9 & 190.6949 &   2.6595 &                  3.60 & J1939-6342 &   J1256-0547 \\
		SN2020ue & 2020-01-19 23:14:52.5 $-$ 2020-01-20 03:46:49.6 & 190.6949 &   2.6595 &                  3.61 & J1939-6342 &   J1256-0547 \\
		SN2020ue & 2020-02-06 00:55:23.8 $-$ 2020-02-06 05:28:24.9 & 190.6949 &   2.6595 &                  3.62 & J1939-6342 &   J1256-0547 \\
		SN2020hvf & 2020-04-24 15:44:50.4 $-$ 2020-04-24 21:22:13.9 & 170.3602 &   3.0147 &                  4.98 & J0408-6545 &   J1058+0133 \\
		SN2020hvf &  2020-05-01 17:15:5.2 $-$ 2020-05-01 22:50:12.7 & 170.3602 &   3.0147 &                  4.97 & J0408-6545 &   J1058+0133 \\
		SN2021smj & 2021-07-10 12:45:57.1 $-$ 2021-07-10 18:51:27.9 & 186.6940 &   8.8827 &                  4.95 & J0408-6545 &   J1150-0023 \\
		SN2021smj & 2021-07-23 11:50:54.9 $-$ 2021-07-23 17:57:13.6 & 186.6940 &   8.8827 &                  4.95 & J0408-6545 &   J1150-0023 \\
		SN2021qvv &   2021-08-09 11:07:9.8 $-$ 2021-08-09 15:48:2.6 & 187.0122 &   9.8056 &                  3.96 & J0408-6545 &   J1150-0023 \\
		SN2021smj &  2021-09-05 08:41:42.0 $-$ 2021-09-05 14:48:8.7 & 186.6940 &   8.8827 &                  4.95 & J0408-6545 &   J1150-0023 \\
		SN2022ffv &  2022-03-31 09:22:5.0 $-$ 2022-03-31 15:01:52.3 &  54.1238 & -35.2893 &                  4.97 & J0408-6545 &   J0440-4333 \\
		SN2022ffv & 2022-04-01 09:29:45.1 $-$ 2022-04-01 17:21:37.1 &  54.1238 & -35.2893 &                  4.97 & J0408-6545 &   J0440-4333 \\
		SN2022ffv &  2022-04-14 08:22:1.1 $-$ 2022-04-14 14:01:48.5 &  54.1238 & -35.2893 &                  4.98 & J0408-6545 &   J0440-4333 \\
		GRB220730A &  2022-08-01 18:36:8.7 $-$ 2022-08-01 22:40:46.5 & 225.0143 & -69.4959 &                  3.46 & J1939-6342 &   J1619-8418 \\
		GRB220730A & 2022-08-03 16:30:46.1 $-$ 2022-08-03 20:36:27.8 & 225.0143 & -69.4959 &                  3.47 & J1939-6342 &   J1619-8418 \\
		GRB220730A & 2022-08-10 15:15:52.6 $-$ 2022-08-10 19:22:14.4 & 225.0143 & -69.4959 &                  3.47 & J1939-6342 &   J1619-8418 \\
		GRB200522A &  2022-08-16 23:22:42.4 $-$ 2022-08-17 03:23:4.3 &   5.6820 &  -0.2832 &                  3.47 & J1939-6342 &   J0022+0014 \\
		GRB200907B & 2022-08-19 05:18:50.2 $-$ 2022-08-19 09:23:44.0 &  89.0290 &   6.9062 &                  3.47 & J0408-6545 &   J0521+1638 \\
		GRB210919A &   2022-08-20 05:17:2.6 $-$ 2022-08-20 09:19:8.5 &  80.2545 &   1.3115 &                  3.46 & J0408-6545 &   J0503+0203 \\
		GRB210726A &  2022-08-21 11:26:1.1 $-$ 2022-08-21 15:29:43.0 & 193.2909 &  19.1875 &                  3.46 & J1331+3030 &   J1330+2509 \\
		GRB200411A & 2022-08-22 02:16:28.7 $-$ 2022-08-22 06:03:30.9 &  47.6641 & -52.3176 &                  3.22 & J0408-6545 &   J0210-5101 \\
		GRB210323A & 2022-08-23 20:17:47.4 $-$ 2022-08-24 00:23:13.2 & 317.9461 &  25.3699 &                  3.47 & J1939-6342 &   J2236+2828 \\
		GRB200219A & 2022-08-24 00:47:39.5 $-$ 2022-08-24 04:52:25.3 & 342.6385 & -59.1196 &                  3.46 & J0408-6545 &   J2329-4730 \\
		GRB220730A & 2022-09-08 15:37:10.1 $-$ 2022-09-08 19:42:19.8 & 225.0143 & -69.4959 &                  3.46 & J1939-6342 &   J1619-8418 \\
		SN2020eyj & 2022-05-07 16:10:32.8 $-$ 2022-05-07 21:05:17.3 & 167.9466 &  29.3893 &                  3.96 & J1120+1420 &   J0408-6545 \\
		SN2020eyj &  2022-10-28 04:47:5.0 $-$ 2022-10-28 09:42:45.5 & 167.9466 &  29.3893 &                  3.97 & J1120+1420 &   J0408-6545 \\
		\hline
	\end{tabular}
%\end{table*}
\end{sidewaystable}
Almost all observations were calibrated using version 2.0 of the ProcessMeerKAT pipeline, with the exception of the GRB~220730A observations that were processed using version 1.1 of the pipeline~\citep[{\sc ProcessMeerKAT};][]{2021ursi.confE...4C}. As part of the calibration process, parts of the spectrum known to be heavily contaminated with radio frequency interference (RFI) were flagged, as well as the edges of the bandpass, resulting in a bandwidth of approximately 800 MHz centered at 1.3~GHz, which is a reduction from 856 MHz in the raw data. The flagged frequency ranges differed slightly in some observations and can be over 50\% of the MeerKAT L-band, resulting in the variations in the center frequencies shown in Table~\ref{tab:obstimescales8}. Calibration was performed in parallel over 9 spectral windows (11 for ProcessMeerKAT 1.1). Common Astronomy Software Applications~\citep[{\sc CASA;}][]{2022arXiv221002276T} tools were used to perform complex gain and flux calibration, along with flagging for RFI, using the tasks \textit{tfcrop} and \textit{rflag}. After two rounds of calibration and flagging, the spectral windows were recombined into a single measurement set for imaging.

Imaging was performed using tclean, with parameters to account for non-coplanar baselines. The w-project gridder was used with 128 planes, the gain was set to 0.08, and multi-term multi-frequency synthesis was used with 2 Taylor terms and scales of 0, 5, and 15. The imaging process started with a shallow image, using a threshold of 1 mJy. This image was used to create a model for phase self-calibration. Additional RFI flagging was performed using rflag. After self-calibration, images with an integration time of approximately 30 minutes were created by combining adjacent 15 minute scans into a single image. Primary beam correction was performed on the 30-minute images using the katbeam library \citep{2022AJ....163..135D}. The mean rms noise and the number of images for the 30-minute timescale is summarized in Table~\ref{tab:obstimescales30}. The center frequency, number of images, and median rms noise for the 8-second images is summarized in Table~\ref{tab:obstimescales8}. The 8-second images of the GRB 200522A field were not included due to artifacts from bright field sources.

\begin{table}
	\caption{Number of 8-second images for each field, center frequency, and the median RMS noise.}
	\label{tab:obstimescales8}
	\begin{tabular}{l|l|l|l}
		\hline
		Target & Center Freq. & No. of Images & Median RMS Noise  \\
		& (MHz) &  & ($\mu$Jy) \\
		\hline
		SN2019np & 1098 & 2799 & 318\\
		SN2019np & 1498 & 2799 & 195\\
		SN2019muj & 1098 & 1860 & 241\\ 
		SN2019muj & 1498 & 1860 & 144\\ 
		SN2020ue & 1098 & 4831 & 272\\
		SN2020ue & 1498 & 4831 & 164\\
		SN2020hvf & 1098 & 4489 & 262\\
		SN2020hvf & 1498 & 4489 & 143 \\
		SN2021smj & 1098 & 6716 & 365 \\
		SN2021smj & 1498 & 6716 & 160 \\
		SN2021qvv & 1098 & 1792 & 523 \\
		SN2021qvv & 1498 & 1792 & 286 \\
		SN2022ffv & 1098 & 7611 & 482 \\
		SN2022ffv & 1498 & 7612 & 283 \\
		GRB220730A & 1030 & 6265 & 268 \\
		GRB220730A & 1480 & 6265 & 139 \\
		GRB200907B & 1030 & 1568 & 303 \\
		GRB200907B & 1480 & 1568 & 168 \\
		GRB210919A & 1030 & 1565 & 741 \\
		GRB210919A & 1480 & 1565 & 277\\
		GRB210726A & 1030 & 1566 & 311\\
		GRB210726A & 1480 & 1566 & 151\\
		GRB200411A & 1030 & 1454 & 282\\
		GRB200411A & 1480 & 1454 & 126\\
		GRB210323A & 1030 & 1567 & 245\\
		GRB210323A & 1480 & 1567 & 136\\
		GRB200219A & 1030 & 1564 & 272 \\
		GRB200219A & 1480 & 1564 & 135 \\
		SN2020eyj & 1134 & 3573 & 353\\
		SN2020eyj & 1498 & 3573 & 171\\
	\end{tabular}
\end{table}
\begin{table}
	\caption{Number of 30-minute images for each field and the mean RMS noise.}
	\label{tab:obstimescales30}
	\begin{tabular}{l|l|l}
		\hline
		Target & Number of Images & Mean RMS Noise ($\mu$Jy) \\
		\hline
		SN2019np & 12 & 31\\
		SN2019muj & 12 & 31\\
		SN2020ue & 33 & 37\\
		SN2020hvf & 10 & 20\\
		SN2021smj & 30 & 44\\
		SN2021qvv & 8 & 77\\
		SN2022ffv & 17 & 38\\
		GRB220730A & 28 & 23\\
		GRB200522A & 7 & 140\\
		GRB200907B & 7 & 47\\
		GRB210919A & 7 & 96\\
		GRB210726A & 7 & 33\\
		GRB200411A & 6 & 19\\
		GRB210323A & 7 & 27\\
		GRB200219A & 7 & 21\\
		SN2020eyj & 16 & 36\\
	\end{tabular}
\end{table}

\section{Methods}
\label{sec:methods3}
% \subsection{Sourcefinding and Association}

We used the LOFAR Transients Pipeline~\citep[{\sc TraP};][]{2015A&C....11...25S} for source finding, source association, and image quality control. TraP is also designed to calculate variability metrics based on source light curves, but we chose to do these calculations separately in order to account for a 10\% systematic error. Quality control is taken care of in TraP through multiple checks. The rms noise is measured in the inner part of the image by removing the brightest pixels. The measured noise in the image is compared to global allowed minimum and maximum values and put into a running histogram. After the first 100 images, if the noise is greater than 3 sigma from the mean, it is rejected. This resulted in the rejection of six images. There are other reasons that an image could have been rejected, such as beam shape, for which we relaxed the ellipticity requirements to a factor of 10, and with that, these six images were the only rejected ones. 

Before determining variability, we only selected sources that were within 0.8 degrees of the center of the field. Given the fall-off of sensitivity and complex shape of the beam, this restriction prevents effects on the flux of our sources in ways that are difficult to model accurately. Additionally, the noise within this distance to the center of the primary beam does not appreciably increase. In order to assess the variability of the sources, we use the variability measures V and $\eta$ as defined in \citet{2015A&C....11...25S}, where $I$ is the flux measurement of a source, $\xi$ is the average flux weighted by the inverse of the flux measurement errors $\sigma$, and averages are indicated by hyphens above the quantities in these equations:
\begin{equation}\label{Veqn}
	V_{\nu} = \frac{1}{\bar{I}_{\nu}}\sqrt{\frac{N}{N-1} (\bar{I_{\nu}^2} - \bar{I_{\nu}}^2)}
\end{equation}
\begin{equation}\label{etaeqn}
	\eta =  \frac{1}{N-1} \sum_{i=1}^{N} \frac{(I_{\nu,i} - \xi_{I_{\nu}})^2}{\sigma_{\nu,i}^2}   
\end{equation} 
We recomputed V and $\eta$ for every source found by TraP, incorporating a 10\% systematic error into the calculations of $\eta$, by adding it in quadrature to the flux errors reported by TraP. For the 8-second images we took the additional step of computing an $\eta$ value in 14 images at a time for each source and keeping the largest computed $\eta$ value. This number of images was selected by testing the maximum number of images that can be used before a bright, less than or equal to 8 second transient, gets washed out into an insignificant amount of variability. In other words, this step ensures that any short timescale variation does not get averaged out over the large number of images included in the 8-second dataset. For both the 8-second and 30-minute images, we considered sources with an $\eta > 2$ to be a candidate variable. This value was chosen in our first study \citep{commensal1} by examining a large number of sources and examining their variability.  We also determined a signal-to-noise detection threshold in the same way as \citet{2022MNRAS.517.2894R} and \citet{commensal1}. We took a sample of 50\% of the pixels that were included in the transient search and determined the threshold at which less than 1 false transient detection would occur by assuming the pixels follow a Gaussian distribution. Using this method, we determined a threshold of approximately 5.4 times the rms noise in the image for the 30-minute images, and 6.4 times the rms noise in the image for the 8-second images. This threshold is only a rough approximation since the pixels do not follow a Gaussian distribution; see also the discussion in \citet{2022MNRAS.517.2894R}. %Figure~\ref{fig:pixels} and 

In order to reduce the number of sidelobes flagged as potential transients, we rejected all but the brightest source in a circular region of 5 beam widths around every source (measured by the beam's major axis), following \citet{commensal1}. Some sidelobes still remained after this step, and we could have made this radius larger to attempt to catch more, but we did not want to risk losing any more potential candidates, so we removed these sidelobe sources in later steps by visual determination.

% \begin{figure}
	%     % \includegraphics[width=0.9\textwidth]{ro_figure_1.png}
	% 	\includegraphics[width=\columnwidth]{pixels.png}
	%     \caption{Sample of pixels used for determining the detection threshold along with the fitted Gaussian.}
	%     \label{fig:pixels}
	% \end{figure}

\section{Results}
\label{sec:results3}
\subsection{30-Minute images}
After performing the aforementioned steps to find transients, and then systematically reducing the number of spurious sources, we were left with 874 sources that had an $\eta$ value greater than 2 or with a detection in only a single image. We made animations of these sources\footnote{Code available at https://github.com/dentalfloss1/sharedscripts as FindOutliers.py} to further categorize them. After visual examination, 289 sources were identified as sidelobes of brighter sources, 57 sources were due to noise patterns, 477 sources could not be categorized due to the lack of contrast in the visualization, 28 sources failed to make a proper animation due to only being detected in a single image, and 24 sources appeared to be astrophysical sources. Before remaking animations of the 477 sources that needed improved contrast, a forced fit was done at the source locations of all of these sources in addition to the sources that were only detected in a single image and those that appeared to be astrophysical sources. This forced-fitting step was performed to ensure that any potential variables or transients were not due the effects of a low signal-to-noise source. After this step, only 52 sources remained with an $\eta$ value above two. Each of the 52 sources were examined, and sources that appeared to have variability due to issues with the image, such as an elliptical beam shape rotating over time to cause two sources to merge into one apparently bright source, or an extremely bright source causing issues across the image, were rejected, leaving 14 sources. 

Out of these 14 sources, 11 showed variability on timescales longer than the observation length. Of the remaining three sources, the sources in the 30-minute images show variability within the $4-5$~hour observation, which could have been due to a bright source in the field. In order to determine whether this bright source was affecting these sources, we re-ran TraP on the full observations, measuring the corrected $\eta$ value for each of the three sources. After this step, the corrected $\eta$ value for one of the sources was less than two and this source was removed from consideration. Therefore, our final count of variables in this data set is 13.

\subsection{8-Second images}
\label{sec:8sectransients}
In contrast to~\citet{commensal1}, we split the bandwidth of the MeerKAT L-band in half in attempt to increase the likelihood of capturing potential steep spectrum or narrowband transients. We performed the same steps to find transients and remove spurious transients as in the 30-minute images, resulting in 1803 sources with $\eta$ greater than two. Movies were made of each source, and after removing sources that were either artifacts, likely satellites or other RFI, or had variability that seemed to be non-astrophysical in origin, we were left with one transient source.

The detected source, source 96178, is detected at a right ascension and declination of ($317.565\pm0.003$, $25.397\pm0.003$) degrees (J2000), and it has a duration of 8 to 16 seconds. This source was initially only detected in the lower frequencies of the observing band in the GRB 210323A field. Upon detecting this transient, we re-calibrated and re-imaged the observation with the cross hand polarizations included in order to obtain Stokes V polarization images. We found that the source seemed to have significant stokes V polarization in the lowest frequencies. We additionally split the bandwidth into four instead of two parts and found that the observed flux to be highest at the lowest frequencies. A summary of the measurements is given in Table~\ref{tab:src96178measurements}. Similar to the 30 minute images, the primary beam correction was then performed using the \citep[\textsc{katbeam};][]{2022AJ....163..135D} library.

\label{tab:src96178measurements}
\begin{table}
	\caption{Primary beam corrected flux measurements of Stokes I and V flux of source 96178 using \textsc{TraP}.}
	\begin{tabular}{llll}
		\hline
		Center Freq. & Bandwidth & $F_{int,I}$ & $F_{int,V}$ \\
		(MHz) & (MHz) & (mJy) & (mJy) \\
		\hline
		989.0 & 200.0 & $1.93\pm0.59$ & $-1.44\pm0.19$ \\
		1207.1 & 200.0 & $1.18\pm0.68$ & $-0.43\pm0.24$ \\
		1388.9 & 200.0 & $1.03\pm0.48$ & $0.35\pm0.16$  \\
		1570.8 & 200.0 & $0.96\pm0.57$ & $0.05\pm0.19$ \\
		\hline
	\end{tabular}
\end{table}

\section{Discussion}
\label{sec:discussion3}

\subsection{Variable Sources}
In the transient searches of the 30-minute images, no transients were found, but we did identify 13 variable sources. These variable sources have already been verified to be of an astrophysical nature and not due to any calibration or statistical issues. In order to investigate the sources further, an initial search of the source positions were compared against catalogs contained in the Vizier \citep[][]{vizier} catalog access tool using a search radius of 2 arcseconds from the position of the variable source reported by \textsc{TraP}. This initial search yielded an unrealistically large number of optical, infrared, and near-infrared counterparts. Therefore, we limited counterparts to those detected by TESS \citep{2015JATIS...1a4003R}, X-ray catalogs, gamma-ray catalogs and radio catalogs. See Section~\ref{sec:multiwavelengthclass} for more information. The Rapid ASKAP Continuum Survey (RACS) \citep[][]{2020PASA...37...48M} and the VLA Sky Survey (VLASS) \citep[][]{2020PASP..132c5001L} survey data were also searched for counterparts to the variable sources.  For some sources, catalog data was already available, for other sources PySE \citep[][]{2018ascl.soft05026S} was used to fit a source in the survey data at the variable source location. Cutouts from VLASS were obtained from Canadian Initiative for Radio Astronomy Data Analysis (CIRADA, cutouts.cirada.ca) while cutouts from RACS were obtained from the CSIRO Data Portal CASDA Cutout Service (https://data.csiro.au).

\subsubsection{Sources with Radio Counterparts}

\begin{figure*}
	\includegraphics[width=0.48\textwidth]{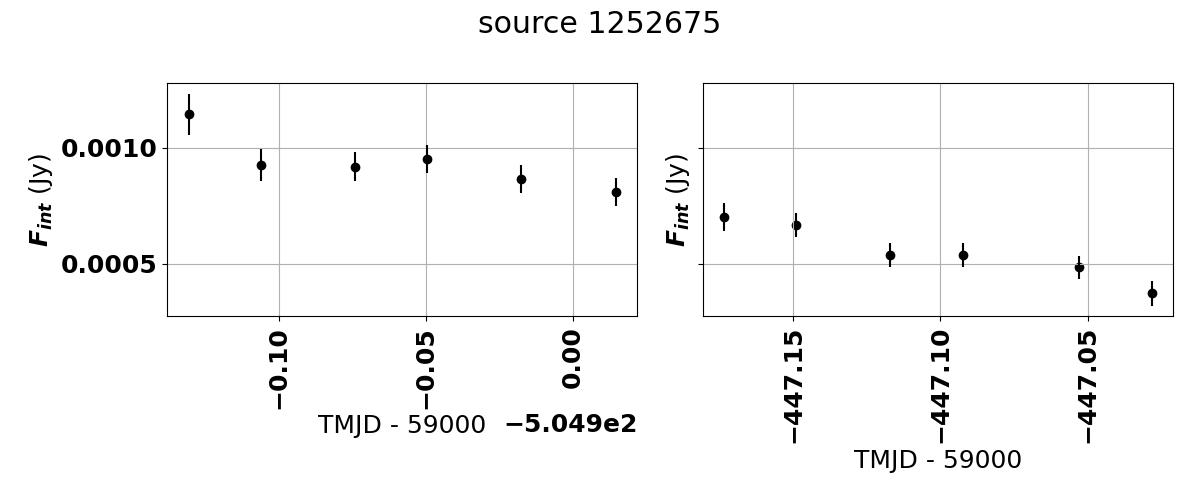}
	\includegraphics[width=0.48\textwidth]{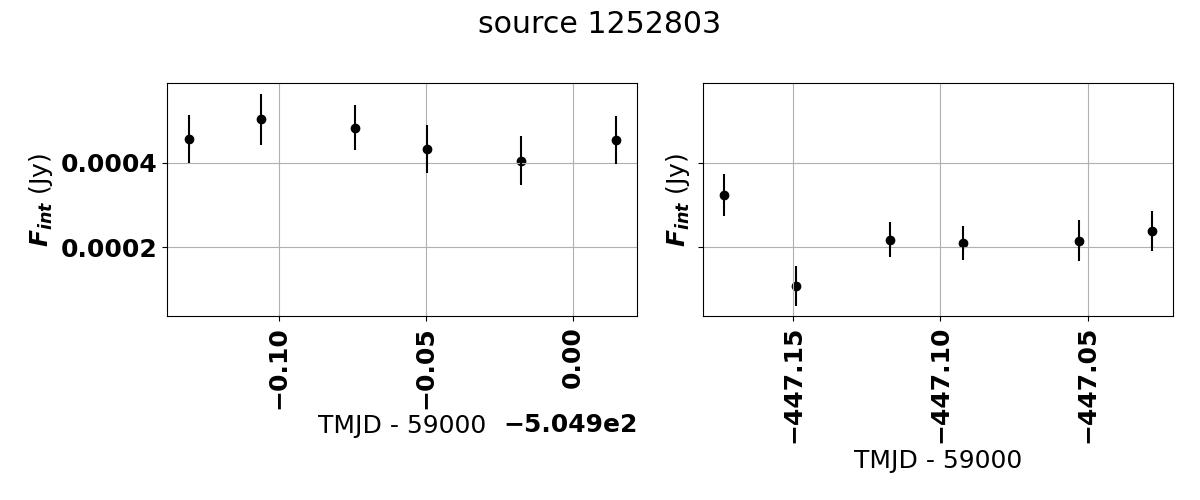}
	\includegraphics[width=0.48\textwidth]{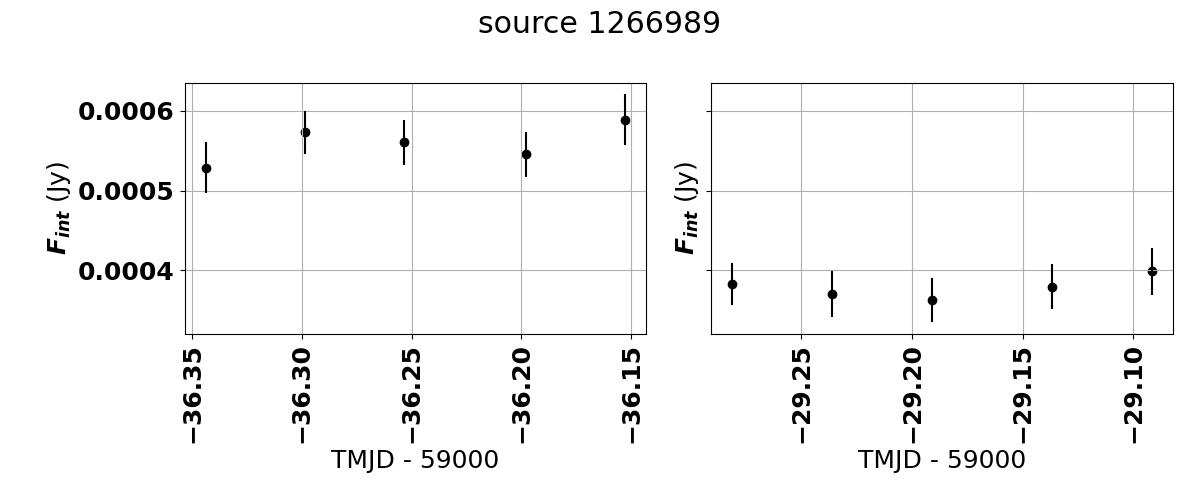}
	\includegraphics[width=0.48\textwidth]{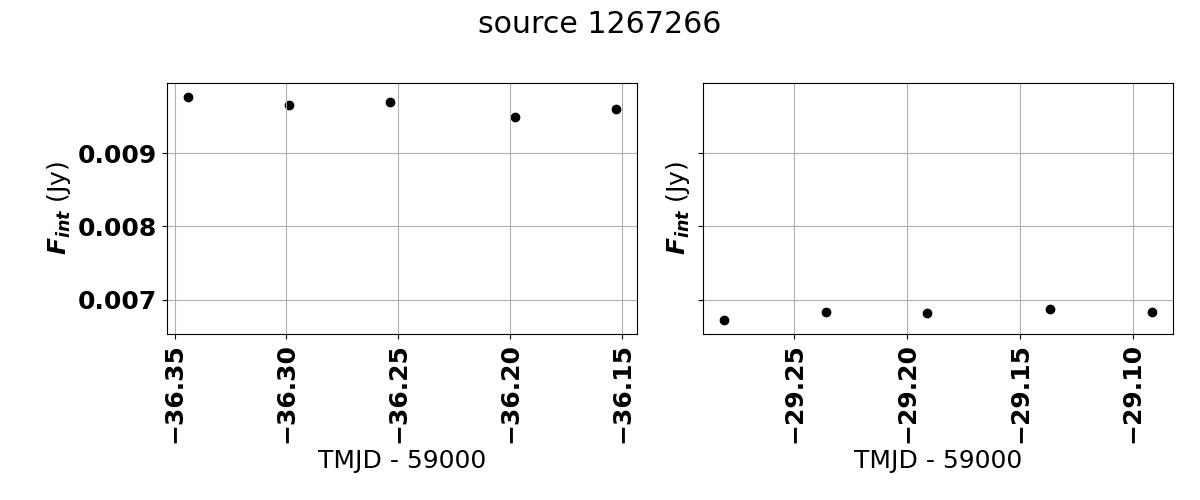}
	\includegraphics[width=\textwidth]{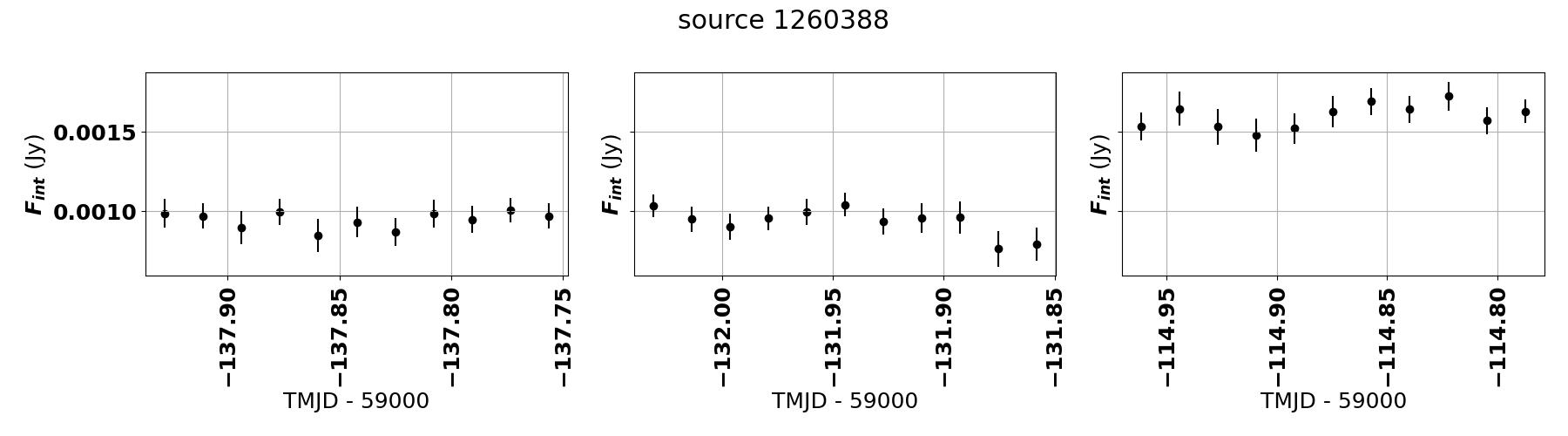}

	\includegraphics[width=\textwidth]{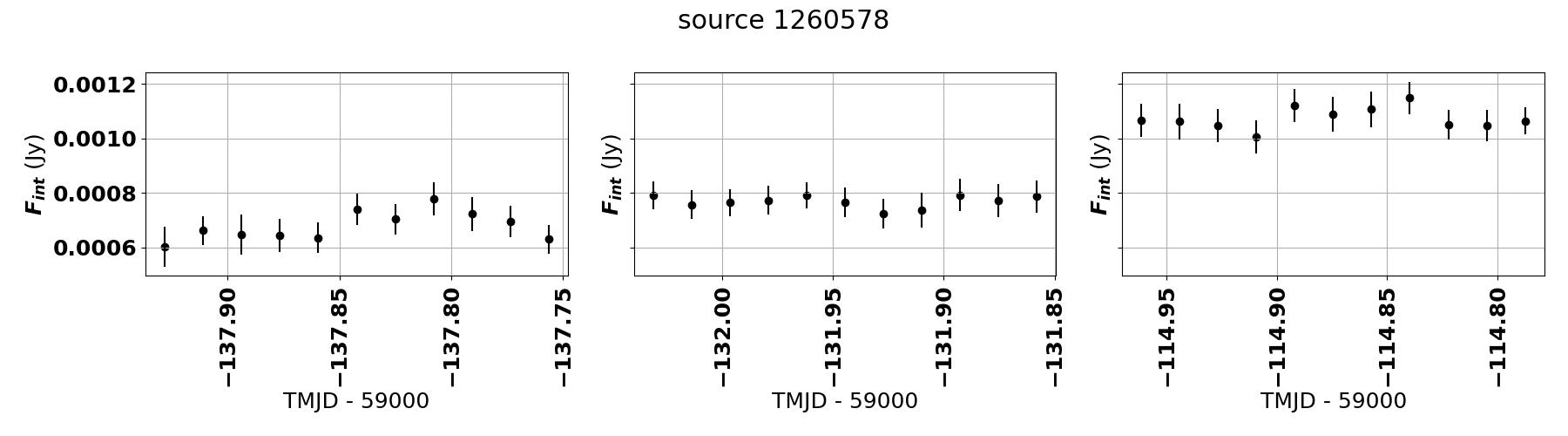}
	\caption{Light curves for the variable sources, not including source 1290071.}
	\label{fig:varsrc1}
\end{figure*}

All of our variable sources together with their counterpart measurements, or forced measurements, are shown in Tables~\ref{tab:radiocounterparts1} and~\ref{tab:radiocounterparts2}. The light curves for these sources are shown in Figures~\ref{fig:varsrc1}-\ref{fig:src1290071multi}. 

Our measurements from MeerKAT at 1.3~GHz agree fairly well with the catalog measurements for most of these sources, with the exceptions being the two sources discussed specifically in this section. There is one notably bright source, source 1267266, which showed variability between the NVSS and FIRST surveys. The integrated flux was 10 mJy in NVSS and 7.2 mJy in FIRST. In Figure~\ref{fig:varsrc1}, we see this same range of variability in the MeerKAT data as well.

\subsubsection{Source 1278651}
\begin{figure*}

	\includegraphics[width=0.8\textwidth]{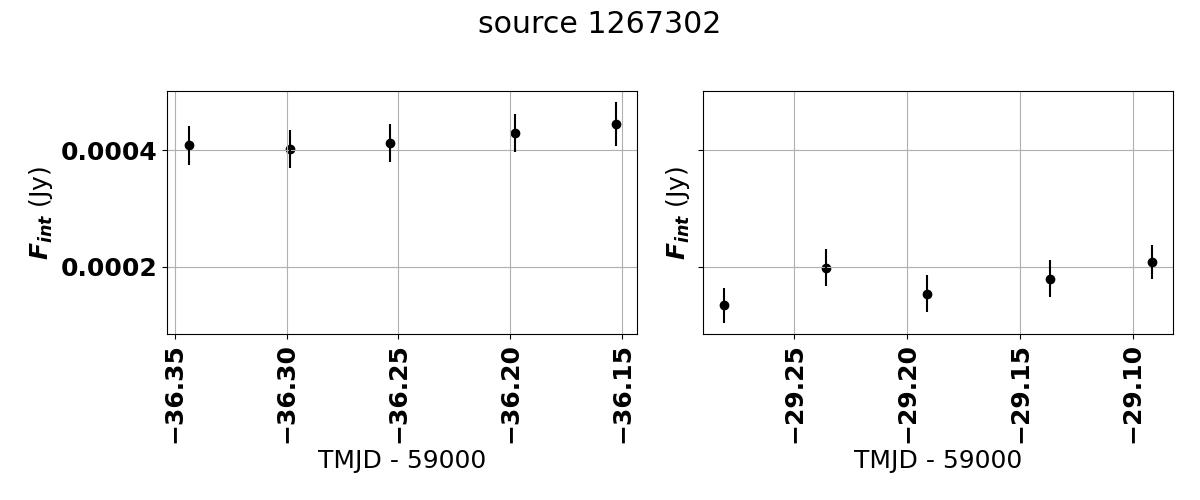}
	\includegraphics[width=0.33\textwidth]{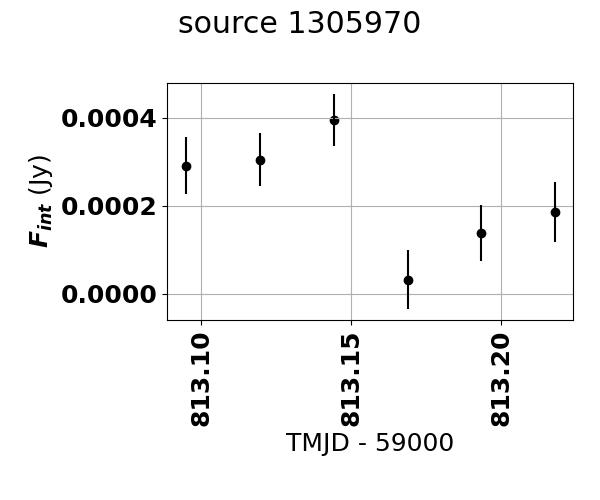}
	\includegraphics[width=0.33\textwidth]{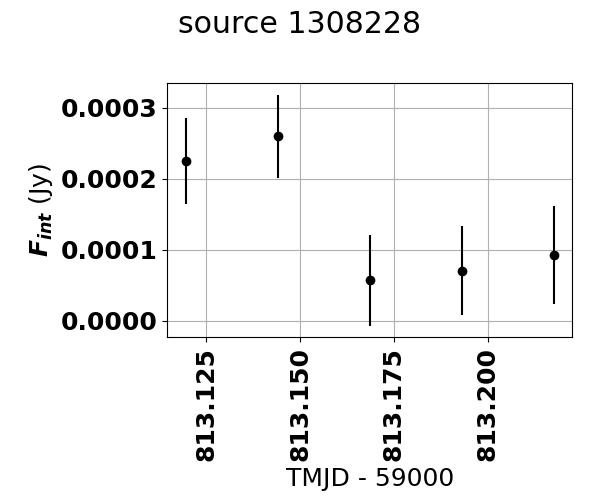}
	\includegraphics[width=0.33\textwidth]{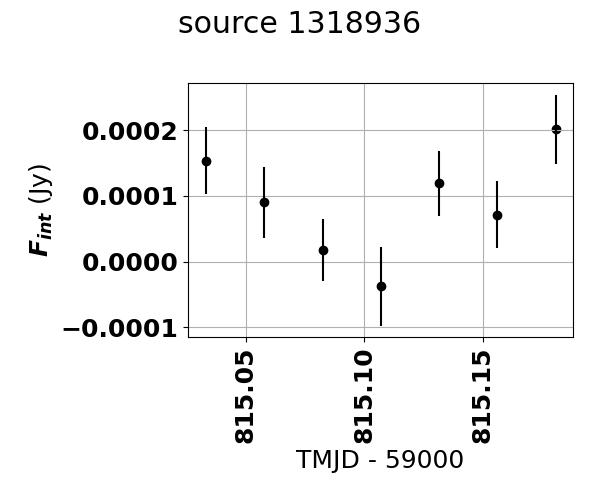}
	\includegraphics[width=\textwidth]{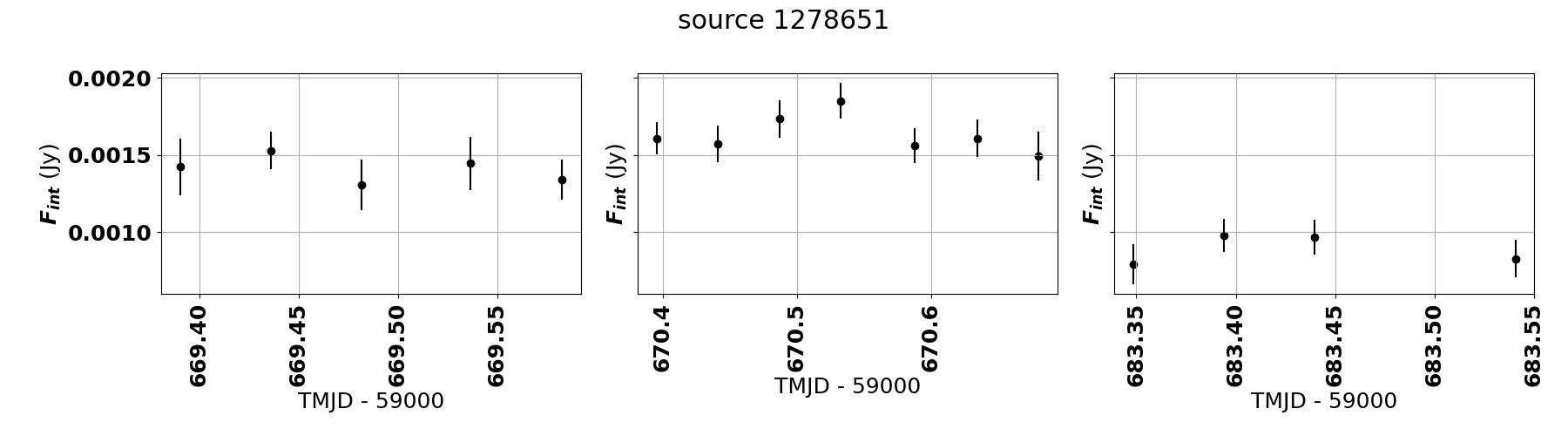}
	\includegraphics[width=\textwidth]{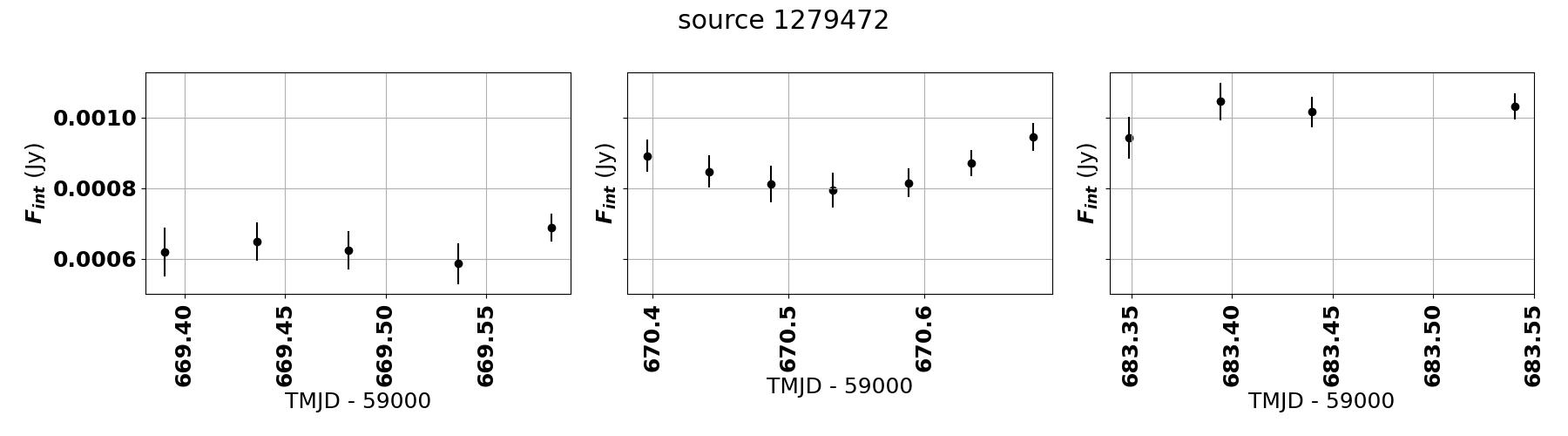}
	
	\caption{Light curves, continued, for the variable sources, not including source 1290071.}
	\label{fig:varsrc2}
\end{figure*}

Source 1278651 shows variability in RACS and VLASS, however the detections are only marginal. The measurements using PySE shown in Table \ref{tab:radiocounterparts2} roughly fall within the range of our measurements at 1.3~GHz (see also Figure~\ref{fig:varsrc2}). This variability in the time between observations could either be intrinsic, due to the signal-to-noise, or due to scintillation as will be discussed in section~\ref{sec:InterstellarScintillation}. 

\begin{figure}
	\centering
	\includegraphics[width=0.4\textwidth]{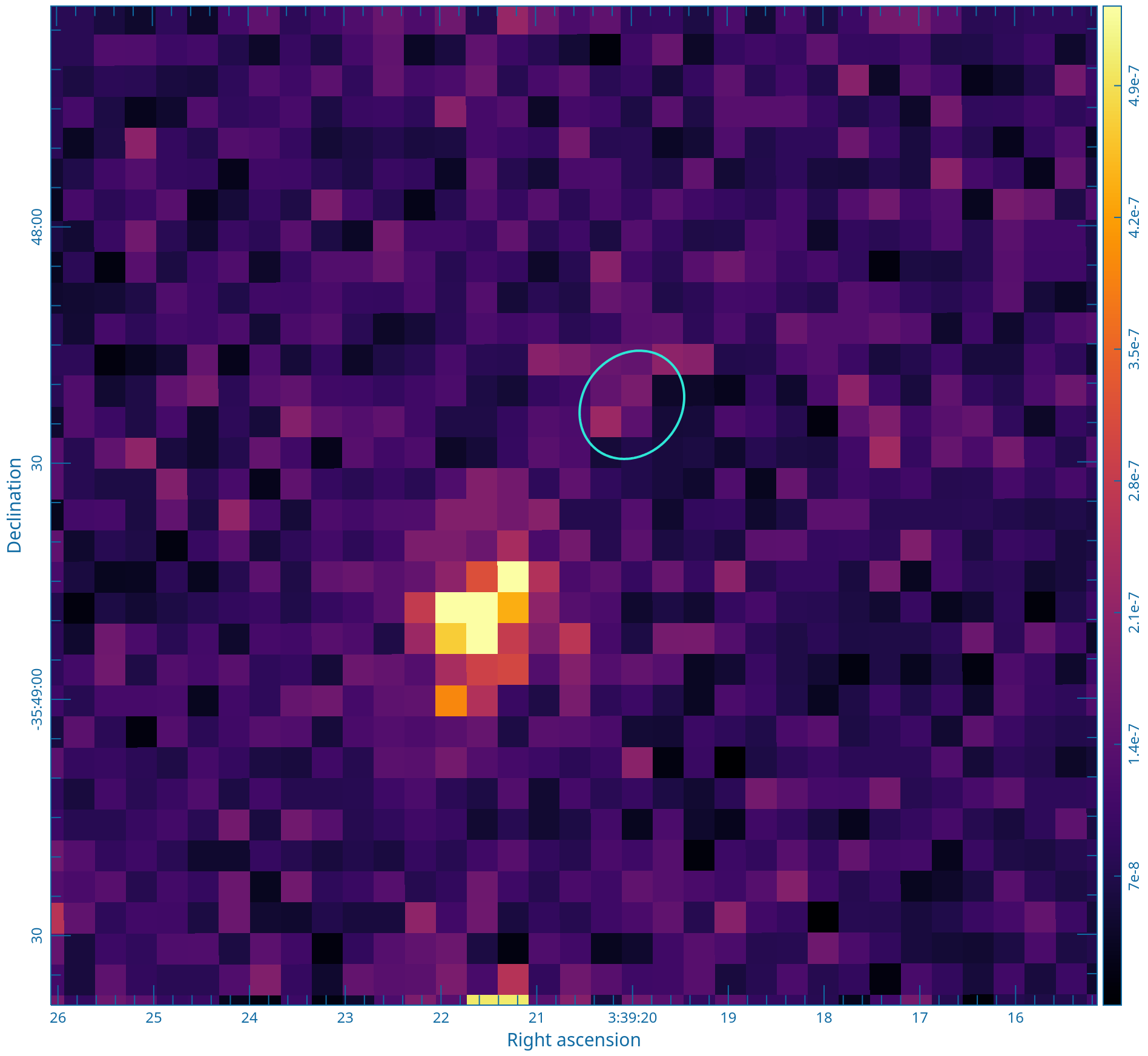}
	\includegraphics[width=0.4\textwidth]{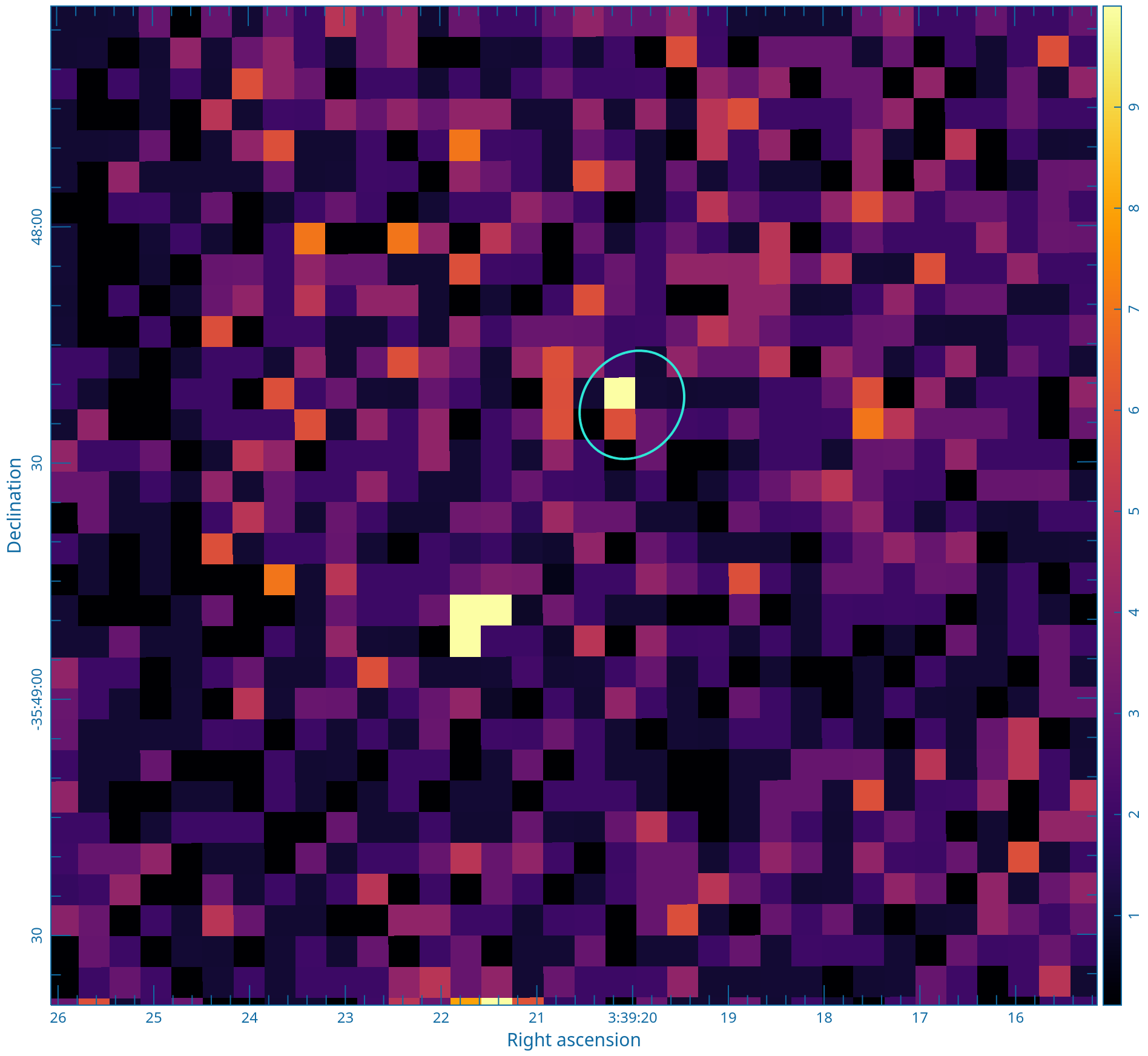}
	\caption{On the left is a co-added image of \textit{Chandra} ACIS data shown with the MeerKAT beam from one of the observations overplotted as an ellipse. On the right is an image from \dataset [Chandra ObsId 4175]{https://doi.org/10.25574/04175} in which the source appears to be detected.}
	\label{fig:chandra}
\end{figure}

This source has conflicting multi-wavelength catalog data: \citet{2021AJ....161..147B} show a star approximately 1.3~kpc away within 1 arcsecond of the source using Gaia~\citep{prusti2016gaia}. However, there is also a faint X-ray source at this location: \citet{2011ApJS..192...10L} detect a source with \textit{Chandra}, reporting a distance of 12 Mpc, which is also present in the second \textit{Chandra} Source Catalog~\citep{2010ApJS..189...37E}. Since this was found in an extragalactic field and the variability is consistent with interstellar scintillation, these X-ray measurements hint at this source being an AGN. In order to examine this source further, we downloaded archival \textit{Chandra} ACIS \citep{2003SPIE.4851...28G} data using the Chandra Interactive Analysis of Observations \citep[\textsc{CIAO};][]{2006SPIE.6270E..1VF} software tools. The data was reprocessed and merged into a single image shown in Figure~\ref{fig:chandra} along with an observation in which it was detected. Based on this image, the \textit{Chandra} source does not appear to be confidently detected in the combined image. However, in one of the observations, the source is clearly detected. This source could possibly be variable in X-rays. The image was made from \dataset [Chandra ObsId 4175]{https://doi.org/10.25574/04175},\dataset [4176]{https://doi.org/10.25574/04176},\dataset [3949]{https://doi.org/10.25574/03949},\dataset [2942]{https://doi.org/10.25574/02942},\dataset [17548]{https://doi.org/10.25574/17548},\dataset [17541]{https://doi.org/10.25574/17541},\dataset [17540]{https://doi.org/10.25574/17540},\dataset [16234]{https://doi.org/10.25574/16234},\dataset [16233]{https://doi.org/10.25574/16233},\dataset [16232]{https://doi.org/10.25574/16232}, and \dataset [16231]{https://doi.org/10.25574/16231}.

\subsubsection{Source 1290071}
\begin{figure*}
	\includegraphics[width=\textwidth]{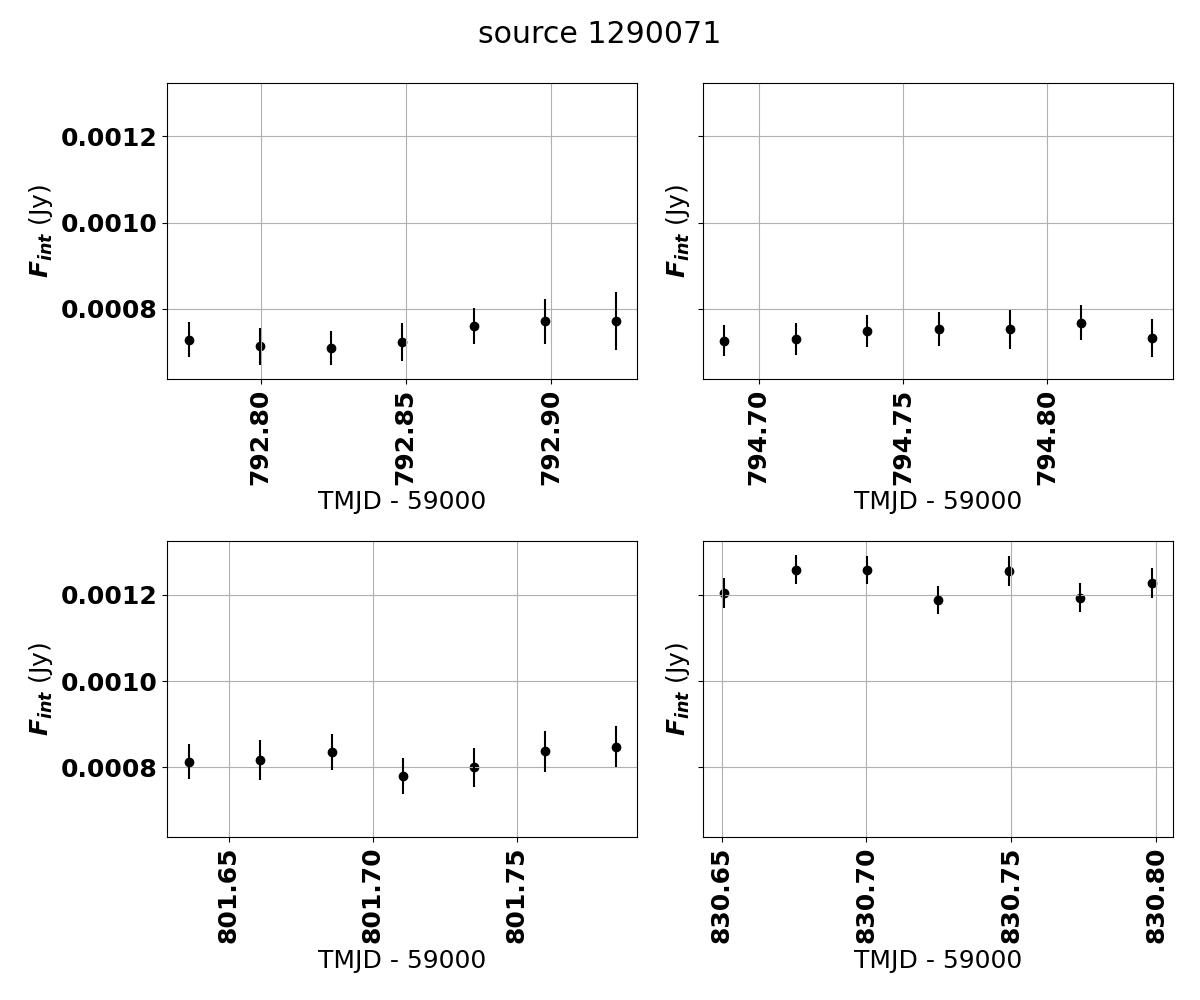}
	\caption{Light curve showing all observations of source number 1290071, which shows variability on the observation-to-observation level. It also has radio counterparts and the variability timescale of around 30 days does not match the refractive scintillation timescale of around 60 days}.
	\label{fig:src1290071multi}
\end{figure*}

Source 1290071 is notable for its lack of variability in RACS. Our measurements show this source varying between a lower flux of around 0.8 mJy and a higher flux state of around 1.2 mJy (see Figure~\ref{fig:src1290071multi}). Notably, the RACS measurements at both 890 MHz and 1400 MHz are consistent with our higher flux measurements and stable in flux. However, these measurements do not necessarily rule out interstellar scintillation as the main cause of variability, due to the cadence of the RACS observations.

\begin{table*}[htb!]
	\caption{Radio catalog counterparts for variable sources in our MeerKAT survey that are visible to both VLASS and RACS. When indicated, source measurements were made using PySE \citep{2018ascl.soft05026S} on image cutouts acquired from the Canadian Initiative for Radio Astronomy Data Analysis (CIRADA) and the CSIRO Data Portal CASDA Cutout Service. VLASS data is from \citet{2021ApJS..255...30G}, NVSS data from \citet{1998AJ....115.1693C}, FIRST data from \citet{2015ApJ...801...26H}, and RACS data are from \citet{RACS1} and \citet{RACS2}.}  
	\label{tab:radiocounterparts1}
	\begin{tabular}{lllllllll}
		\hline
		Source &       RA &      Dec &  $F_{int}$  &    $F_{pk}$&   Frequency &                Date &       Survey & Measurement \\
		&        &       &    (mJy)  &    (mJy) &   (GHz) &    &   &                  \\
		\hline
		1252675 & 157.3791 & 29.8533 & $2.5\pm0.5$ & $1.4\pm0.2$ & 2.7 & 2019-04-22 23:38:59 & VLASS & catalog \\
		1252675 & 157.3791 & 29.8533 & $1.2\pm0.5$ & $1.1\pm0.2$ & 1.4 & 2021-01-03 19:03:30 & RACS & PySE \\
		1252675 & 157.3791 & 29.8533 & $1.3\pm0.5$ & $1.2\pm0.3$ & 1.4 & 2021-01-02 19:10:06 & RACS & PySE \\
		\hline
		1252803 & 157.2391 & 29.9353 & $-0.04\pm0.3$ & $0.8\pm4.0$ & 2.7 & 2019-04-22 23:38:59 & VLASS & PySE \\
		1252803 & 157.2391 & 29.9353 & $-0.08\pm0.2$ & $0.07\pm0.1$ & 2.7 & 2021-12-03 9:39:52 & VLASS & PySE \\
		1252803 & 157.2391 & 29.9353 & $0.06\pm0.3$ & $-0.5\pm0.2$ & 1.4 & 2021-01-03 19:03:30 & RACS & PySE \\
		1252803 & 157.2391 & 29.9353 & $0.07\pm0.4$ & $-0.6\pm2.0$ & 1.4 & 2021-01-02 19:10:06 & RACS & PySE \\
		\hline
		1260388 & 191.0687 & 2.334 & $1.6\pm0.3$ & $1.3\pm0.1$ & 2.7 & 2019-04-21 8:11:40 & VLASS & catalog \\
		1260388 & 191.0687 & 2.334 & $1.7\pm1.1$ & $1.5\pm0.5$ & 0.89 & 2020-04-30 14:06:00 & RACS & PySE \\
		1260388 & 191.0687 & 2.334 & $1.7\pm1.1$ & $1.5\pm0.5$ & 0.89 & 2020-04-30 14:23:00 & RACS & PySE \\
		1260388 & 191.0687 & 2.334 & $1.7\pm1.1$ & $1.5\pm0.5$ & 0.89 & 2020-05-01 13:24:00 & RACS & PySE \\
		1260388 & 191.0687 & 2.334 & $1.7\pm1.1$ & $1.5\pm0.5$ & 0.89 & 2020-05-01 13:41:00 & RACS & PySE \\
		1260388 & 191.0687 & 2.334 & $1.4\pm0.4$ & $1.4\pm0.2$ & 1.4 & 2021-01-03 22:22:34 & RACS & PySE \\
		1260388 & 191.0687 & 2.334 & $1.3\pm0.4$ & $1.3\pm0.2$ & 1.4 & 2021-01-09 22:37:42 & RACS & PySE \\
		1260388 & 191.0687 & 2.334 & $1.4\pm0.4$ & $1.3\pm0.2$ & 1.4 & 2021-01-03 22:02:39 & RACS & PySE \\
		1260388 & 191.0687 & 2.334 & $1.3\pm0.4$ & $1.3\pm0.2$ & 1.4 & 2022-07-30 3:24:18 & RACS & PySE \\
		\hline
		1260578 & 190.8379 & 2.4131 & $1.5\pm0.3$ & $1.2\pm0.1$ & 2.7 & 2019-04-21 8:11:58 & VLASS & catalog \\
		1260578 & 190.8379 & 2.4131 & $-0.09\pm0.7$ & $1.8\pm8.0$ & 0.89 & 2020-04-30 14:06:00 & RACS & PySE \\
		1260578 & 190.8379 & 2.4131 & $-0.1\pm0.7$ & $1.1\pm3.5$ & 0.89 & 2020-04-30 14:23:00 & RACS & PySE \\
		1260578 & 190.8379 & 2.4131 & $-0.1\pm0.7$ & $1.5\pm5.6$ & 0.89 & 2020-05-01 13:24:00 & RACS & PySE \\
		1260578 & 190.8379 & 2.4131 & $-0.01\pm0.7$ & $1.6\pm6.4$ & 0.89 & 2020-05-01 13:41:00 & RACS & PySE \\
		1260578 & 190.8379 & 2.4131 & $0.7\pm0.3$ & $0.7\pm0.2$ & 1.4 & 2021-01-03 22:22:34 & RACS & PySE \\
		1260578 & 190.8379 & 2.4131 & $0.8\pm0.3$ & $0.7\pm0.2$ & 1.4 & 2021-01-03 22:02:39 & RACS & PySE \\
		1260578 & 190.8379 & 2.4131 & $0.6\pm0.4$ & $0.6\pm0.2$ & 1.4 & 2021-01-09 22:37:42 & RACS & PySE \\
		1260578 & 190.8379 & 2.4131 & $0.6\pm0.4$ & $0.5\pm0.2$ & 1.4 & 2021-01-09 22:17:57 & RACS & PySE \\
		\hline
		1266989 & 170.58 & 2.8567 & $0.2\pm0.3$ & $-0.07\pm0.09$ & 2.7 & 2020-08-22 21:59:48 & VLASS & PySE \\
		1266989 & 170.58 & 2.8567 & $0.5\pm0.2$ & $0.4\pm0.1$ & 2.7 & 2023-02-02 12:06:59 & VLASS & PySE \\
		1266989 & 170.58 & 2.8567 & $0.5\pm0.5$ & $0.3\pm0.2$ & 0.89 & 2023-12-31 21:04 & RACS & PySE \\
		1266989 & 170.58 & 2.8567 & $0.5\pm0.5$ & $0.3\pm0.2$ & 0.89 & 2024-01-25 19:12 & RACS & PySE \\
		1266989 & 170.58 & 2.8567 & $-0.09\pm0.3$ & $0.2\pm0.3$ & 1.4 & 2021-01-16 20:45:26 & RACS & PySE \\
		1266989 & 170.58 & 2.8567 & $-0.08\pm0.2$ & $0.2\pm0.4$ & 1.4 & 2021-01-09 20:58:10 & RACS & PySE \\
		\hline
		1267266 & 170.4347 & 2.8388 & $10.0\pm0.5$ & & 1.4 & 1995-02-27 & NVSS & catalog \\
		1267266 & 170.4347 & 2.8388 & $7.2$ & $7.2$ & 1.4 & 1998-07-01 & FIRST & catalog \\
		1267266 & 170.4347 & 2.8388 & $6.1\pm0.2$ & $6.1\pm0.2$ & 2.7 & 2018-01-12 11:12:44 & VLASS & PySE \\
		1267266 & 170.4347 & 2.8388 & $8.8\pm1.4$ & $7.2\pm0.4$ & 0.89 & 2020-05-01 12:50:35 & RACS & PySE \\
		1267266 & 170.4347 & 2.8388 & $5.9\pm0.3$ & $5.9\pm0.2$ & 1.4 & 2021-01-09 20:58:10 & RACS & PySE \\
		1267266 & 170.4347 & 2.8388 & $5.6\pm0.3$ & $5.6\pm0.2$ & 1.4 & 2021-01-16 20:45:26 & RACS & PySE \\
		\hline
		1267302 & 170.4211 & 3.3767 & $0.0\pm0.3$ & $3.2\pm60.0$ & 2.7 & 2020-08-22 21:39:45 & VLASS & PySE \\
		1267302 & 170.4211 & 3.3767 & $-0.02\pm0.2$ & $0.7\pm3.6$ & 2.7 & 2023-02-02 12:40:38 & VLASS & PySE \\
		1267302 & 170.4211 & 3.3767 & $0.1\pm0.6$ & $-0.7\pm1.5$ & 0.89 & 2020-05-01 12:50 & RACS & PySE \\
		1267302 & 170.4211 & 3.3767 & $0.1\pm0.6$ & $-0.7\pm2.0$ & 0.89 & 2020-05-01 13:07 & RACS & PySE \\
		1267302 & 170.4211 & 3.3767 & $-0.3\pm0.3$ & $-0.2\pm0.1$ & 1.4 & 2021-01-16 20:45:26 & RACS & PySE \\
		
		\hline\end{tabular}
\end{table*}
\begin{table*}[htb!]
	\caption{Radio counterparts for variable sources continued from Table~\ref{tab:radiocounterparts1}. Sources in italics are discussed in the text.}
	\label{tab:radiocounterparts2}  
	\begin{tabular}{lllllllll}
		\hline
		Source &       RA &      Dec &  $F_{int}$  &    $F_{pk}$&   Frequency &                Date &       Survey & Measurement \\
		&        &       &    (mJy)  &    (mJy) &   (GHz) &    &   &                  \\
		\hline
		\textit{1278651} & 54.8335 & -35.8063 & $0.34\pm0.27$ & $0.3\pm0.1$ & 2.7 & 2020-10-29 7:01:24 & VLASS & PySE \\
		\textit{1278651} & 54.8335 & -35.8063 & $0.6\pm0.2$ & $0.5\pm0.1$ & 2.7 & 2023-06-13 18:06:24 & VLASS & PySE \\
		\textit{1278651} & 54.8335 & -35.8063 & $0.74\pm0.71$ & $0.5\pm0.3$ & 0.89 & 2019-04-27 5:54:00 & RACS & PySE \\
		\textit{1278651} & 54.8335 & -35.8063 & $0.72\pm0.7$ & $0.5\pm0.3$ & 0.89 & 2019-04-28 3:14:00 & RACS & PySE \\
		\textit{1278651} & 54.8335 & -35.8063 & $1.0\pm0.3$ & $1.0\pm0.2$ & 1.4 & 2020-12-31 13:34:27 & RACS & PySE \\
		\textit{1278651} & 54.8335 & -35.8063 & $1.0\pm0.3$ & $0.9\pm0.2$ & 1.4 & 2021-01-07 13:25:42 & RACS & PySE \\
		\hline
		1279472 & 53.7036 & -35.6255 & $1.2\pm0.3$ & $1.1\pm0.2$ & 2.7 & 2020-10-29 7:01:24 & VLASS & catalog \\
		1279472 & 53.7036 & -35.6255 & $0.5\pm0.4$ & $0.4\pm0.2$ & 0.89 & 2019-04-27 5:54 & RACS & PySE \\
		1279472 & 53.7036 & -35.6255 & $0.5\pm0.4$ & $0.4\pm0.2$ & 0.89 & 2019-04-28 3:14 & RACS & PySE \\
		1279472 & 53.7036 & -35.6255 & $1.0\pm0.3$ & $0.9\pm0.1$ & 1.4 & 2020-12-31 13:34:27 & RACS & PySE \\
		1279472 & 53.7036 & -35.6255 & $1.0\pm0.3$ & $0.9\pm0.1$ & 1.4 & 2020-12-31 13:51:12 & RACS & PySE \\
		\hline
		\textit{1290071} & 224.7278 & -59.5997 & $1.2\pm0.3$ & $1.2\pm0.2$ & 0.89 & 2019-05-07 12:41:05 & RACS & PySE \\
		\textit{1290071} & 224.7278 & -59.5997 & $1.2\pm0.3$ & $1.2\pm0.2$ & 0.89 & 2019-05-07 13:27:32 & RACS & PySE \\
		\textit{1290071} & 224.7278 & -59.5997 & $1.2\pm0.3$ & $1.2\pm0.2$ & 0.89 & 2020-05-02 15:48:10 & RACS & PySE \\
		\textit{1290071} & 224.7278 & -59.5997 & $1.2\pm0.3$ & $1.2\pm0.2$ & 0.89 & 2020-06-19 13:35:51 & RACS & PySE \\
		\textit{1290071} & 224.7278 & -59.5997 & $1.5\pm0.3$ & $1.5\pm0.2$ & 1.4 & 2021-03-14 20:49:38 & RACS & PySE \\
		\textit{1290071} & 224.7278 & -59.5997 & $1.5\pm0.3$ & $1.5\pm0.2$ & 1.4 & 2021-03-15 18:54:55 & RACS & PySE \\
		\textit{1290071} & 224.7278 & -59.5997 & $1.5\pm0.3$ & $1.5\pm0.2$ & 1.4 & 2022-03-04 19:17:23 & RACS & PySE \\
		\hline
		1305970 & 48.3284 & -52.7501 & $-0.1\pm0.4$ & $0.4\pm1.0$ & 0.89 & 2020-03-28 6:00 & RACS & PySE \\
		1305970 & 48.3284 & -52.7501 & $-0.2\pm0.4$ & $0.1\pm0.2$ & 0.89 & 2020-03-28 6:17 & RACS & PySE \\
		1305970 & 48.3284 & -52.7501 & $-0.1\pm0.4$ & $0.3\pm0.5$ & 0.89 & 2020-03-28 6:34 & RACS & PySE \\
		1305970 & 48.3284 & -52.7501 & $0.3\pm0.3$ & $0.2\pm0.1$ & 1.4 & 2021-01-23 9:55:38 & RACS & PySE \\
		1305970 & 48.3284 & -52.7501 & $0.3\pm0.3$ & $0.3\pm0.1$ & 1.4 & 2021-01-24 10:40:35 & RACS & PySE \\
		\hline
		1308228 & 48.6663 & -52.4636 & $0.1\pm0.4$ & $-0.6\pm0.1$ & 0.89 & 2020-03-28 6:00 & RACS & PySE \\
		1308228 & 48.6663 & -52.4636 & $0.1\pm0.4$ & $-0.4\pm0.8$ & 0.89 & 2020-03-28 6:17 & RACS & PySE \\
		1308228 & 48.6663 & -52.4636 & $0.1\pm0.4$ & $-0.4\pm0.8$ & 0.89 & 2020-03-28 6:34 & RACS & PySE \\
		1308228 & 48.6663 & -52.4636 & $0.0\pm0.3$ & $8.0\pm400$ & 1.4 & 2021-01-23 9:55:38 & RACS & PySE \\
		1308228 & 48.6663 & -52.4636 & $-0.1\pm0.3$ & $0.1\pm0.2$ & 1.4 & 2021-01-24 10:40:35 & RACS & PySE \\
		\hline
		1318936 & 341.5495 & -59.0375 & $0.4\pm0.4$ & $0.2\pm0.1$ & 0.89 & 2019-05-07 19:00 & RACS & PySE \\
		1318936 & 341.5495 & -59.0375 & $0.4\pm0.4$ & $0.2\pm0.1$ & 0.89 & 2020-03-29 1:51 & RACS & PySE \\
		1318936 & 341.5495 & -59.0375 & $0.3\pm0.3$ & $0.2\pm0.1$ & 1.4 & 2022-07-30 2:59:58 & RACS & PySE \\
		\hline\end{tabular}
\end{table*}

\subsection{Other Multi-wavelength Data and Classification}
\label{sec:multiwavelengthclass}
An initial search of the available optical catalogs with Vizier \citep[][]{vizier} revealed that, in addition to the aforementioned radio data, almost all of the variable sources, except for 1290071 and 1305970, have an optical counterpart within two arcseconds of the position of the variable source reported by \textsc{TraP}. It is difficult, however, to confirm the association between the optical and radio sources. \cite{2015ApJ...801...26H} discusses the challenge of confirming counterparts between relatively low resolution radio telescopes and optical catalogs, a challenge that is only more difficult with deep optical catalogs. It is notable that the two sources that have no optical counterpart have low declinations where there are much fewer instruments and less catalog data with which to compare. While it is likely that at least some of these sources are scintillating active galactic nuclei (AGN), the coarse localization makes it difficult to confirm the true nature of these sources. Due to these aforementioned challenges, we only considered radio, X-ray, or gamma-ray sources due to the higher potential of association with a radio source, as well as TESS sources since they have lightcurves available to which we can compare time variability.
% \begin{figure*}
	%     % \includegraphics[width=0.9\textwidth]{ro_figure_1.png}
	% 	\includegraphics[width=\textwidth]{myradoptvars.png}
	%     \caption{Radio versus optical flux density for a variety of variable sources, adapted from \citet{2018MNRAS.479.2481S}, with the variable sources from our MeerKAT survey in black squares.}
	%     \label{fig:radoptplot}
	% \end{figure*}

%a flux of $6.9^{1.9}_{2.7}\times10^{-7}\text{ ph}\text{ cm}^{-2}\text{ s}^{-1}$ between 0.5-2 keV and a flux of $1.4^{0.3}_{1.4}\times10^{-7}\text{ ph}\text{ cm}^{-2}\text{ s}^{-1}$ between 2-8 keV. There is also a source at this location in the second Chandra Source Catalog~\citep{2010ApJS..189...37E} with a flux of $7.1\times10^{-7}\text{ ph} \text{ cm}^{-2}\text{ s}^{-1}$. 

\subsection{Interstellar Scintillation}
\label{sec:InterstellarScintillation}
Interstellar scintillation is a common cause of variability in the radio sky. \citet{1998MNRAS.294..307W} provides a review of the phenomenon with examples of its effects. Interstellar scintillation is caused by the radio waves from a distant, typically compact, object passing through the ionized component of the interstellar medium. This interaction causes a variation in the flux of the radio source characterized by the modulation index, $m$. The modulation index is defined by the fractional variation in the flux: $m=\sigma/\mu$, where $\sigma$ is the variation in flux, and $\mu$ is the average flux. This variability can be characterized over a particular timescale called the variability timescale, $t_{var}$. 

In order to predict the expected variability timescale and modulation index for a particular location on the sky, it is necessary to compare the observing frequency with the transition frequency, $\nu_0$. At the transition frequency, the modulation index, or fractional variability is expected to be at its highest, that is of order unity. If the observing frequency is above the transition frequency, the scattering from the ionized interstellar medium is considered ``weak'', if the observing frequency is below the transition frequency it is considered ``strong''. All of our observations occur in the ``strong'' scattering regime. Furthermore, within this regime, variability can either be from diffractive scintillation or from refractive scintillation. Diffractive scintillation occurs on narrow bandwidths and short timescales, while refractive scintillation occurs over longer timescales and over larger bandwidths. On the timescales and observing frequencies probed by our 30-minute images, refractive scintillation is the most dominant type of scintillation. 

Maps of the expected scintillation properties from the interstellar medium can be made if the distribution of charged particles is known. \citet{2019arXiv190708395H} uses $H_{\alpha}$ intensity maps to generate a model of the electron content and distribution of our galaxy. From this model, the accompanying software also computes the expected transition frequency, variability timescale, and modulation index.

We use the software described in~\citet{2019arXiv190708395H} to compute expected values of refractive scintillation at the pointing center of each field. A summary of the aforementioned scintillation parameters for each field are shown in Table~\ref{tab:scinttab}. Several fields have very large modulation indices, which we can compare to $V$, the variability metric calculated for each source. All of the sources we consider variable are shown in Table~\ref{tab:sourcevar}, with the variability statistics and an estimated upper limit on the timescale of variation estimated by visual inspection of the light curves. 

\begin{table*}
	\centering
	\begin{tabular}{llllll}
		\hline
		Target &       RA (deg) &      Dec (deg) &    Modulation   &  Timescale &   $\nu_0$ \\
		&        &       &  Index     &  (days) &  (GHz) \\
		\hline
		SN2019muj &  36.5771 &  -9.8359 & $0.69\pm   0.09$   &  $ 2.0 \pm         0.8$ &  2.5 \\
		SN2020ue & 190.6949 &   2.6595 & $0.53\pm   0.03$   &  $ 4.4 \pm         0.6$ &  3.9 \\
		SN2020hvf & 170.3602 &   3.0147 & $0.66\pm   0.07$   &  $ 2.3 \pm         0.8$ &  2.6 \\
		SN2021smj & 186.6940 &   8.8827 & $0.72\pm   0.11$   &  $ 1.7 \pm         0.8$ &  2.3 \\
		SN2021qvv & 187.0122 &   9.8056 & $0.71\pm   0.10$   &  $ 1.8 \pm         0.7$ &  2.3 \\
		SN2022ffv &  54.1238 & -35.2893 & $0.56\pm   0.13$   &  $ 3.9 \pm         2.7$ &  3.6 \\
		SN2020eyj & 167.9466 &  29.3893 & $0.65\pm   0.07$   &  $ 2.3 \pm         0.7$ &  2.7 \\
		GRB220730A & 225.0143 & -69.4959 & $0.29\pm   0.01$   &  $64.1 \pm         9.5$ & 11.5 \\
		GRB200522A &   5.6820 &  -0.2832 & $0.73\pm   0.12$   &  $ 1.7 \pm         0.8$ &  2.2 \\
		GRB200907B &  89.0290 &   6.9062 & $0.30\pm   0.01$   &  $59.9 \pm         4.9$ & 10.9 \\
		GRB210919A &  80.2545 &   1.3115 & $0.26\pm   0.01$   &  $64.5 \pm         4.2$ & 14.2 \\
		GRB210726A & 193.2909 &  19.1875 & $0.66\pm   0.07$   &  $ 2.2 \pm         0.6$ &  2.7 \\
		GRB200411A &  47.6641 & -52.3176 & $0.91\pm   2.20$   &  $ 0.9 \pm         6.7$ &  1.5 \\
		GRB210323A & 317.9461 &  25.3699 & $0.37\pm   0.01$   &  $24.5 \pm         2.4$ &  7.5 \\
		GRB200219A & 342.6385 & -59.1196 & $0.74\pm   0.65$   &  $ 1.7 \pm         4.6$ &  2.2 \\
		SN2019np & 157.3415 &  29.5107 & $0.66\pm   0.07$   &  $ 2.4 \pm         0.8$ &  2.7 \\
		\hline
	\end{tabular}
	\caption{Scintillation properties at the center of each field, calculated using~\citet{2019arXiv190708395H}.}
	\label{tab:scinttab}
\end{table*}

\begin{table}
	\centering
	\begin{tabular}{lllllllll}
		\hline
		Source &       RA &      Dec &  $\eta$ &    V &  timescale  &    Modulation   &  Timescale &   $\nu_0$ \\
		        &         &          &         &      &               (days) &     Index       &  (days)     &  (GHz)\\
		\hline
		1252675 & 157.3791 &  29.8533 & 6.03 & 0.12 &      $<58$  & $0.66\pm   0.07$   &  $ 2.4 \pm         0.8$ &  2.7 \\
		1252803 & 157.2391 &  29.9353 & 4.92 & 0.40 &      $<58$  & $0.66\pm   0.07$   &  $ 2.4 \pm         0.8$ &  2.7 \\
		1260388 & 191.0687 &   2.3340 & 3.70 & 0.28 &      $<17$ & $0.53\pm   0.03$   &  $ 4.4 \pm         0.6$ &  3.9\\
		1260578 & 190.8379 &   2.4131 & 2.53 & 0.21 &      $<17$& $0.53\pm   0.03$   &  $ 4.4 \pm         0.6$ &  3.9 \\
		1266989 & 170.5800 &   2.8567 & 2.98 & 0.21 &       $<7$& $0.66\pm   0.07$   &  $ 2.3 \pm         0.8$ &  2.6  \\
		1267266 & 170.4347 &   2.8388 & 3.19 & 0.18 &       $<7$& $0.66\pm   0.07$   &  $ 2.3 \pm         0.8$ &  2.6  \\
		1267302 & 170.4211 &   3.3767 & 8.53 & 0.44 &      $ <7$& $0.66\pm   0.07$   &  $ 2.3 \pm         0.8$ &  2.6  \\
		\textit{1278651} &  54.8335 & -35.8063 & 3.56 & 0.23 &      $<13$& $0.56\pm   0.13$   &  $ 3.9 \pm         2.7$ &  3.6\\
		1279472 &  53.7036 & -35.6255 & 2.54 & 0.19 &       $<0.8$& $0.56\pm   0.13$   &  $ 3.9 \pm         2.7$ &  3.6 \\
		\textit{1290071} & 224.7278 & -69.5997 & 3.12 & 0.23 &      $<29$ & $0.29\pm   0.01$   &  $64.1 \pm         9.5$ & 11.5\\
		1305970 &  48.3284 & -52.7501 & 3.66 & 0.58 &      $ <0.02$ & $0.91\pm   2.20$   &  $ 0.9 \pm         6.7$ &  1.5  \\
		1308228 &  48.6663 & -52.4636 & 3.66 & 0.67 &       $<0.02$ & $0.91\pm   2.20$   &  $ 0.9 \pm         6.7$ &  1.5  \\
		1318936 & 341.5495 & -59.0375 & 2.06 & 0.13 &       $<0.06$ & $0.74\pm   0.65$   &  $ 1.7 \pm         4.6$ &  2.2\\
		\hline
	\end{tabular}
	\caption{All variable sources in our MeerKAT survey with the variability statistics and the estimated timescale of variability. Sources in italics are discussed in the text.}
	\label{tab:sourcevar}
\end{table}

Comparing Tables \ref{tab:scinttab} and~\ref{tab:sourcevar} shows that all but one source can be considered consistent with refractive scintillation. The upper limit on the timescale of variation in source 1290071 is about half of the expected scintillation timescale with the variability parameter, $V$, being close to the modulation index, $m$. It could be that the assumed distance to the scattering screen used in the calculation of the scintillation characteristics is much less, perhaps half of the assumed 1.5~kpc. This change would result in the timescale matching the estimated upper limit of the variability timescale, with the modulation index being consistent as well. However, in Tables~\ref{tab:radiocounterparts1} and~\ref{tab:radiocounterparts2} we see that the radio measurements at 890 MHz of the RACS counterpart are quite stable. The stability of the flux in these measurements is interesting, although the observing cadence is not fine enough to rule out scintillation for this source. 

Source 1278651 is only marginally detected in RACS at 1.4 GHz and is also only marginally detected in one of the VLASS epochs at 2.7 GHz. It is not detected in RACS at 890 MHz, which should have a sensitivity similar to RACS at 1.4 GHz, around 250 $\mu$Jy. We can see from Table~\ref{tab:scinttab} that the modulation index for this source is quite high and the scintillation timescale is relatively short. Therefore these measurements from RACS and VLASS can still be reconciled with our measurements if it is due to refractive scintillation. 

\subsection{Source 96178: A Highly Circularly Polarized Short Transient}

\begin{figure}
	\includegraphics[width=\textwidth]{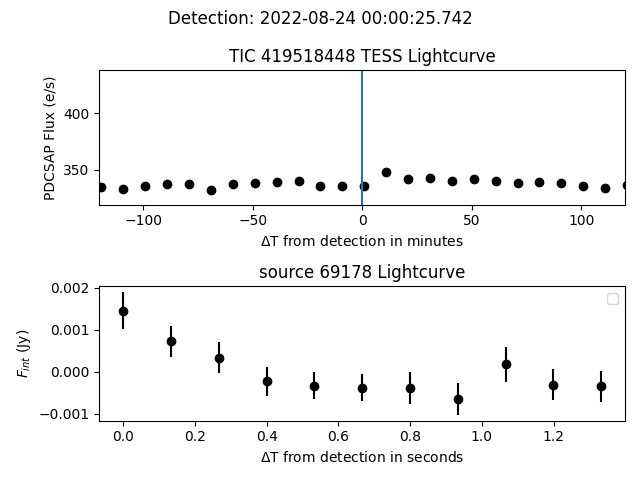}
	\caption{The top plot shows the TESS light curve for TIC 419518448, a potential counterpart to source 96178. The bottom plot shows the MeerKAT light curve. }
	\label{fig:src96178}
\end{figure}

As mentioned in section~\ref{sec:8sectransients}, one source was confidently detected in the lower half of the MeerKAT L-band with approximately 100\% stokes V polarization in the lowest frequencies on a timescale of 8 to 16 seconds. Given that the non-detection in the upper half of the MeerKAT L-band, the emission would need to either be narrowband or have a spectral index less than -1.4. One possible origin of emission of this kind is coherent emission from a stellar flare. It is shorter than typical stellar flares, however due to the near-threshold detection this could be due to only capturing the very peak of the emission. Comparing this source to those in~\citet{2024arXiv240407418D}, the transient duration is significantly shorter than the duration of all of the observations included in the catalog. This transient is approximately 2 arcseconds away from a cataloged TESS M-dwarf star that is 99 pc from Earth (TIC 419518448) \citep{2019AJ....158..138S}. Both the TESS and MeerKAT light curves are shown in Figure~\ref{fig:src96178}. The TESS lightcurve data is available at \dataset[doi:10.17909/s4cr-6z50]{https://dx.doi.org/10.17909/s4cr-6z50}.  These sources could be related, but the association is difficult to confirm without a better localization. It is possible that the slight bump in the TESS light curve could correspond to the MeerKAT transient, however the increase in flux is not very significant. Furthermore, the relatively coarse 10 minute cadence is difficult to compare with the 8 second cadence of the MeerKAT light curve. There were no X-ray, gamma-ray or radio counterparts.

Another possible source type with high stokes V polarization and a steep spectrum would be a pulsar. However, since this source is a transient source, the emission would have to be a giant pulse from a pulsar. There are no pulsars listed in the ATNF Pulsar Catalog \citep{2004IAUS..218..139H} within a quarter of a degree of the source location, making such an origin unlikely. Compared to giant pulses from pulsars, coherent emission from a flare star is expected to be much more common \cite[for a review, see][]{2008arXiv0801.2573O}. The timescale of the emission, potentially up to 16 seconds, also fits well with what would be expected from a flaring star compared to a pulsar, where, at L-band frequencies, the timescales would be expected to be much shorter. Another source type worth considering is the recently discovered long period repeating transients (LPRTs) mentioned in Section~\ref{sec:introduction}. If this transient were a periodic source, the period would most likely be greater than 4 hours, the length of the observations. Such a period would be a bit long even for these LPRTs. Furthermore, these LPRTs are typically found in the galactic plane, while this source is in an extragalactic field. It is nevertheless, still a possibility given the low detection statistics at this time. Since this transient is likely a stellar flare, follow-up was not performed. 

\subsection{Transient Rate Limits}

Using the survey parameters along with the number of detected transients, which is zero on the longer timescales, we can place limits on the transient rate using the Radio Transients Simulations\footnote{https://github.com/dentalfloss1/RaTS} \citep[\textsc{RaTS};][]{2022ascl.soft04007C}. The details of the \textsc{RaTS} code and how it computes the transient rate limits are described in \citet{2022A&C....4000629C}. In summary, the transient rates were computed by simulating transients, aggregating detection statistics, calculating transient rates for each field in the survey by using Poisson statistics, and taking an average over the transient rates weighted by the survey duration of each field and field of view. 

\begin{figure*}
	\includegraphics[width=\textwidth]{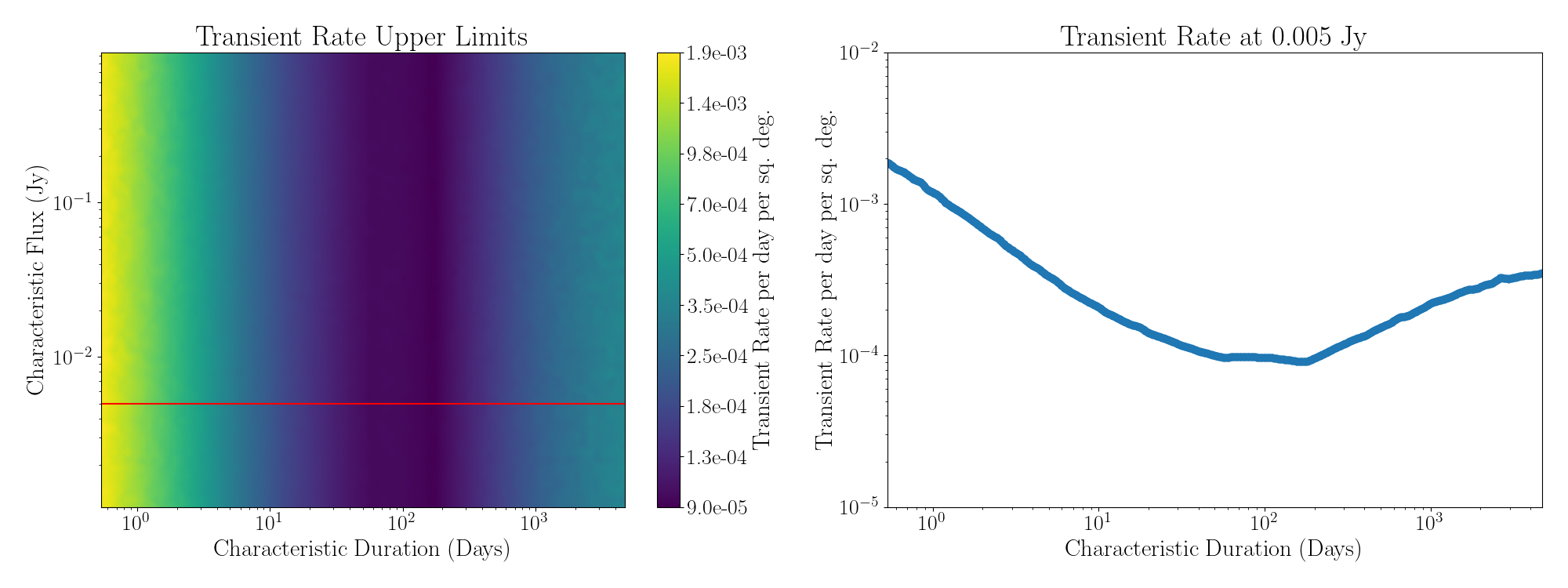}
	\caption{The left panel shows a surface plot with the transient rate upper limit calculated for a top hat light curve on the color axis, and the transient duration and peak flux on the horizontal and vertical axis, respectively. The right panel shows the upper limit on the transient rate as a function of transient duration for transients with a peak flux of 5 mJy. }
	\label{fig:transientrate}
\end{figure*}

Figure~\ref{fig:transientrate} shows the upper limit on transient rates calculated from \textsc{RaTS} for the 30 minute timescale using a simulated transient with a tophat light curve, that is, a lightcurve that is, essentially a step function. The left panel shows the simulated transient duration on the horizontal axis, the simulated transient peak flux on the vertical axis, and the transient rate upper limit on the color axis. In this plot, the darker colors show lower, more constraining upper limits. The red horizontal line marks 5 mJy. The right panel shows the profile of the transient rate upper limit as a function of transient duration for transients with a peak flux of 5 mJy. Our strictest limits show an upper limit of $10^{-4}$ transients per day per square degree for transients with a duration of about 200 days and flux of 5 mJy. Compared to the limits we found in~\citet{commensal1}, this limit is approximately a factor of two lower at a similar transient duration. This may be due to this survey having a larger number of hours on target, 113 versus 100, or, perhaps more likely, having higher quality images on average. 

\begin{figure*}
	\includegraphics[width=\textwidth]{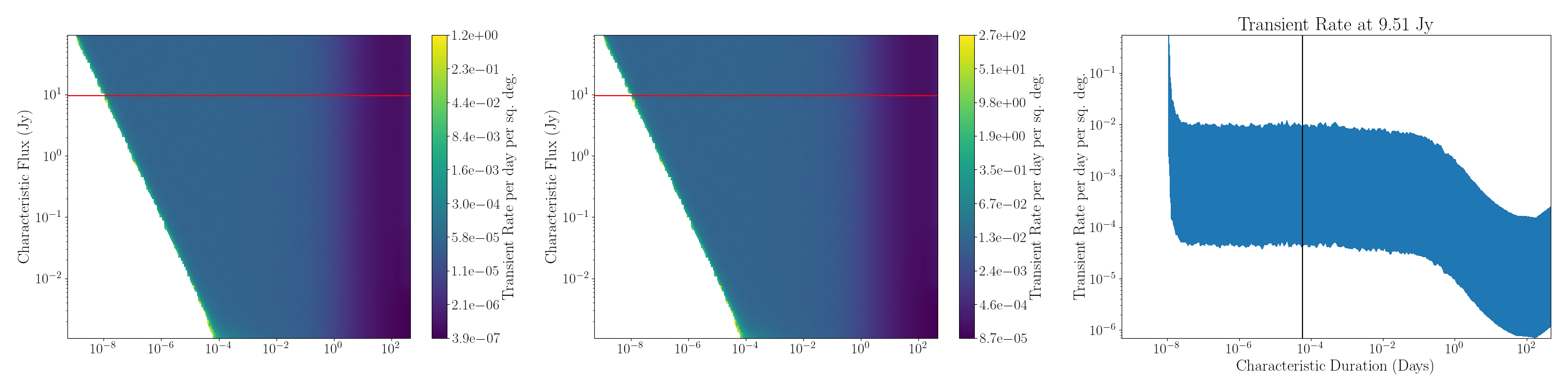}
	\caption{The left panel shows a surface plot with the transient rate lower limit calculated for a top hat light curve on the color axis, and the transient duration and peak flux on the horizontal and vertical axis, respectively. The middle panel is similar to the left panel, but showing the upper limits on the transient rate instead of the lower limits. The right panel shows the lower and upper limits on the transient rate as a function of transient duration for transients with a peak flux of 9.51 Jy. A vertical line is plotted at a duration of 8 seconds.}
	\label{fig:transientrate8sec}
\end{figure*}

Figure~\ref{fig:transientrate8sec} shows the transient rate calculated for the lower half of the L-band at the 8 second timescale. This plot is similar to Figure~\ref{fig:transientrate}, but with an additional plot showing the lower limits on the transient rate. The left panel shows the lower limit on the transient rate on the color axis, the middle panel shows the upper limit on the transient rate on the color axis, and the right panel shows the allowed values for the transient rate in the shaded region as a function of transient duration for a transient with a flux of 9.51 Jy. From the plot in the right panel, we can see that for a transient with a peak flux of 9.51 Jy and on timescales between 8 seconds down to milliseconds, we find a transient rate between $4\times10^{-4}$ and $10^{-2}$ transients per day per square degree. These upper and lower limits are for this particular observing setup and highlights the importance of considering observing bandwidth when computing transient rates. The transient we saw was only visible when we split up the L-band into two different sub-bands. Future updates to the transient rate simulations could include exploring the effects of observing bandwidth, transient spectral width, and dispersion on transient rate computations. Such computations could inform future surveys about the optimal configuration for detecting steep spectrum or narrowband sources.

\section{Looking Forward}
One of the remaining challenges for future radio surveys is accurately characterizing the transient rates or limits on the transient rates for the different kinds of transients that are being detected. The naive, straightforward approach is to simply count up the number of detections, take the timescale to be the duration of the survey, then use a Poisson distribution to compute a simple transient rate. However, as discussed in \cite{2016MNRAS.459.3161C,2017MNRAS.465.4106C,2022A&C....4000629C} this approach frequently produces results that are inaccurate representations of the transient rate due to effects such as transients falling in the gaps between observations and varying sensitivity between observations. Both \cite{commensal1} and this work attempt to more accurately compute the transient rate over a wide part of parameter space using the transient simulations described in \cite{2022A&C....4000629C}. In order to gain a meaningful understanding and comparison between the transient rates from different kinds of sources and surveys, current and future radio surveys will require more rigorous computation of the expected transient rates using techniques such as Monte-Carlo simulations. 

Another significant challenge is the large computational and time cost of performing a transient search such as the one described in this work. One contributing factor for the significant time cost was variability due to imperfect calibration of the images including an imperfect primary beam shape that imparts artificial variability that can be seen in the source lightcurves. Another contributing factor was the choice to include images with elliptical beams. Future surveys with much larger datasets may choose to only select images with nearly circular beams and perhaps raise detection thresholds to avoid sidelobes from bright sources. 

Addressing the large computational requirement may require new techniques for imaging the large amount of data in the survey. \cite{2024MNRAS.531.4805D} describe a technique that accelerates the process by using sky model source subtraction to avoid the computationally intensive cleaning and primary beam correction process. For dense arrays, such as the MeerKAT core or the future DSA-2000, a GPU-based approach that does not perform any cleaning, like that used by the Orville Wideband Imager on the Long Wavelength Array (LWA), could be appropriate. For details on this imager and some applications, see \cite{2021JGRA..12629296V}. If the computational and time cost challenges can be overcome, this survey shows that searching for transients on an 8-second timescale could produce interesting results that otherwise would go undetected.

\section{Conclusion}
\label{sec:conclusion3}

We have conducted a commensal search on short GRB and SN fields using methodology developed in~\citet{commensal1}. In the 30 minute images, we found 13 sources of astrophysical origin with statistically significant variability: source 1252675, 1252803, 1260388, 1260578, 1266989, 1267266, 1267302, 1278651, 1279472, 1290071, 1305970, 1308228, and 1318936. We compared these sources with the RACS and VLASS surveys as well as catalog data from the \textit{Chandra} space telescope. All but source 1290071 are consistent with interstellar scintillation. Many of the sources are likely AGN, but one needs better localization confirming their nature by matching the observations to optical and near-Infrared observations. The variability of the remaining source could be intrinsic in nature. However, it could also be that the distance to the scattering screen used in calculating the expected refractive scintillation is about half of what is estimated. 

On shorter, sub-minute timescales we found one possibly narrowband or steep spectrum transient source with a duration around 8 to 16 seconds and a high degree of circular polarization in the lowest frequencies. This source is roughly consistent with a cataloged M-dwarf star in the TESS Input Catalog \citep{2019AJ....158..138S}, but this association is difficult to confirm without a better localization. Since this source is likely to be a stellar flare, a dedicated follow-up campaign is not currently planned. However, this search does suggest that commensal searches like the current work could be a viable way of finding similar transients, especially if circular polarization is also examined. Finally, we place upper and lower limits on the transient rate using this survey, and find that the upper limits we place on the transient rate on longer timescales is a factor of two better than the commensal search presented in \citet{commensal1}, likely due to more time spent on target and better overall quality of the images.

\section{Acknowledgments}

We would like to thank the referee for providing helpful feedback.

The MeerKAT telescope is operated by the South African Radio Astronomy Observatory (SARAO), which is a facility of the National Research Foundation, an agency of the Department of Science and Innovation. We would like to thank the operators, SARAO staff and ThunderKAT Large Survey Project team.

This work was carried out using the data processing pipelines developed at the Inter-University Institute for Data Intensive Astronomy (IDIA) and available at https://idia-pipelines.github.io. IDIA is a partnership of the University of Cape Town, the University of Pretoria and the University of the Western Cape. We also acknowledge the computing resources provided on the High Performance Computing Cluster operated by Research Technology Services at the George Washington University.

This work made use of the CARTA (Cube Analysis and Rendering Tool for Astronomy) software (DOI 10.5281/zenodo.3377984 –  https://cartavis.github.io).

This research has made use of the CIRADA cutout service at URL cutouts.cirada.ca, operated by the Canadian Initiative for Radio Astronomy Data Analysis (CIRADA). CIRADA is funded by a grant from the Canada Foundation for Innovation 2017 Innovation Fund (Project 35999), as well as by the Provinces of Ontario, British Columbia, Alberta, Manitoba and Quebec, in collaboration with the National Research Council of Canada, the US National Radio Astronomy Observatory and Australia’s Commonwealth Scientific and Industrial Research Organisation.

The ASKAP radio telescope is part of the Australia Telescope National Facility which is managed by Australia’s national science agency, CSIRO. Operation of ASKAP is funded by the Australian Government with support from the National Collaborative Research Infrastructure Strategy. ASKAP uses the resources of the Pawsey Supercomputing Research Centre. Establishment of ASKAP, the Murchison Radio-astronomy Observatory and the Pawsey Supercomputing Research Centre are initiatives of the Australian Government, with support from the Government of Western Australia and the Science and Industry Endowment Fund. We acknowledge the Wajarri Yamatji people as the traditional owners of the Observatory site. This paper includes archived data obtained through the CSIRO ASKAP Science Data Archive, CASDA (https://data.csiro.au).

This research has made use of the VizieR catalogue access tool, CDS, Strasbourg, France (DOI: 10.26093/cds/vizier). The original description of the VizieR service was published in 2000, A\&AS 143, 23.

Some of the data presented in this paper were obtained from the Mikulski Archive for Space Telescopes (MAST) at the Space Telescope Science Institute. The specific observations analyzed can be accessed via \dataset[https://doi.org/10.17909/s4cr-6z50]{https://doi.org/10.17909/s4cr-6z50}. STScI is operated by the Association of Universities for Research in Astronomy, Inc., under NASA contract NAS5–26555. Support to MAST for these data is provided by the NASA Office of Space Science via grant NAG5–7584 and by other grants and contracts.

This research has made use of software provided by the \textit{Chandra} X-ray
Center (CXC) in the application packages CIAO, ChIPS, and Sherpa.

A.H. is grateful for the support by the the United States-Israel Binational Science Foundation (BSF grant 2020203) and by the Sir Zelman Cowen Universities Fund. This research was supported by the ISRAEL SCIENCE FOUNDATION (grant No. 1679/23). This research was supported in part by grant NSF PHY-2309135 to the Kavli Institute for Theoretical Physics (KITP).

MV acknowledges financial support from the Inter-University Institute for Data Intensive Astronomy (IDIA), a partnership of the University of Cape Town, the University of Pretoria and the University of the Western Cape, and from the South African Department of Science and Innovation's National Research Foundation under the ISARP RADIOMAP Joint Research Scheme (DSI-NRF Grant Number 150551) and the CPRR HIPPO Project (DSI-NRF Grant Number SRUG22031677).

%%%%%%%%%%%%%%%%%%%%%%%%%%%%%%%%%%%%%%%%%%%%%%%%%%
\section*{Data Availability}

Data is available upon request by email to sarchast@ttu.edu.

% \begin{figure*}
	% \includegraphics[width=\textwidth]{LEGACY/src713717lc4.png}
	% \caption{On the left is the light curve plotted with linear scaling. On the right, it is plotted on a log-log plot with time measured from the start of the GRB trigger that the image is targeting. }
	% \label{fig:src715717lc4.png}
	% \end{figure*}

\bibliography{draft}{}
\bibliographystyle{aasjournal}

\end{document}